\documentclass{emulateapj}

\shorttitle{Kinematics of DIG in NGC 5775}
\shortauthors{Heald et al.}

\begin{document}

\title{Imaging Fabry-Perot Spectroscopy of NGC 5775: Kinematics of the Diffuse Ionized Gas Halo}
\author{George H. Heald and Richard J. Rand}
\affil{University of New Mexico, Department of Physics and Astronomy, 800 Yale Boulevard NE, Albuquerque, NM 87131}
\author{Robert A. Benjamin}
\affil{University of Wisconsin -- Whitewater, Department of Physics, 800 West Main Street, Whitewater, WI 53190}
\author{Joseph A. Collins}
\affil{University of Colorado, Center for Astrophysics and Space Astronomy, Campus Box 389, Boulder, CO 80309}
\and
\author{Joss Bland-Hawthorn}
\affil{Anglo-Australian Observatory, P.~O. Box 296, Epping, NSW 2121, Australia}

\begin{abstract}
We present imaging Fabry-Perot observations of H$\alpha$ emission in the nearly edge-on spiral galaxy NGC 5775. We have derived a rotation curve and a radial density profile along the major axis by examining position-velocity (PV) diagrams from the Fabry-Perot data cube as well as a CO 2--1 data cube from the literature. PV diagrams constructed parallel to the major axis are used to examine changes in azimuthal velocity as a function of height above the midplane. The results of this analysis reveal the presence of a vertical gradient in azimuthal velocity. The magnitude of this gradient is approximately 1 km s$^{-1}$ arcsec$^{-1}$, or about 8 km s$^{-1}$ kpc$^{-1}$, though a higher value of the gradient may be appropriate in localized regions of the halo. The evidence for an azimuthal velocity gradient is much stronger for the approaching half of the galaxy, although earlier slit spectra are consistent with a gradient on both sides. There is evidence for an outward radial redistribution of gas in the halo. The form of the rotation curve may also change with height, but this is not certain. We compare these results with those of an entirely ballistic model of a disk-halo flow. The model predicts a vertical gradient in azimuthal velocity which is shallower than the observed gradient, indicating that an additional mechanism is required to further slow the rotation speeds in the halo.
\end{abstract}


\section{Introduction}

Gaseous thick disks are observed in the Milky Way and in some external edge-on spirals. In the Milky Way, the layer of vertically extended ionized emission is known as the Reynolds Layer or the warm ionized medium (WIM). This phase consists of gas with T $\sim 10^4$ K and $n_e\sim 0.1$ cm$^{-3}$, and has a scale height of about 1 kpc \citep{reynolds93}.

In external galaxies, these diffuse ionized gas (DIG) layers are observed to have widely varying morphologies \citep[e.g.,][]{rkh90,dettmar90,hwr99,mv03}. Not all edge-on spirals show detectable extraplanar (i.e., above the layer of \ion{H}{2} regions) DIG (EDIG) emission. The prominence of a galaxy's EDIG layer is now known to correlate with tracers of star formation in the disk, such as the surface density of FIR emission, the star formation rate as determined by H$\alpha$ luminosity, and the dust temperature \citep[e.g.,][]{rand96,rd00,rd03a,rd03b}. Other gas phases, as well as dust absorption, have also been observed high above the midplane \citep[see, e.g.,][]{coll00,lee01,dettmar04c,frat04c}.

These observations are generally interpreted in terms of a star formation driven disk-halo flow such as that described by the chimney \citep{ni89} or galactic fountain \citep[e.g.,][]{sf76,breg80} model, and we adopt that viewpoint in this paper. Rising hot gas may originate in chimneys but warm, swept up ambient gas may also be pushed upwards to significant heights. The evolution of such gas and the interaction of different extraplanar phases is not at all well understood. However, the observations described above strongly suggest that such flows exist.

A likely effect of disk-halo cycling is redistribution of gas in the disk. The mass flux rates estimated thus far \citep[cf.][]{bregman94,wang95,frat02b} imply that this process is responsible for moving a large amount of gas. Whether the cycling process is dominated by ballistic motion, (magneto-) hydrodynamic effects, or both, is still an open question. But the cycling may affect the distribution and rate of star formation in the disk, and may result in extraplanar star formation \citep[e.g.,][]{tull03}. That a disk-halo flow is responsible for some of the observed high-velocity clouds (HVCs) is also a possibility \citep[e.g.,][]{wakker97}. For these reasons, it is important to understand the kinematics of this disk-halo cycling.

Studies of extraplanar H$\,$I emission have so far provided the most spatially complete velocity information about gas which may be participating in disk-halo interactions. Halo gas is seen to lag the underlying disk rotation in NGC 891 \citep{ssv97,frat04c}, NGC 5775 \citep{lee01}, and NGC 2403 \citep{sss00,frat01}. This result is expected in a disk-halo flow model because as gas is lifted into the halo, it feels a weaker gravitational potential, migrates radially outward, and thus its rotation speed drops in order to conserve angular momentum. Since DIG has turned out to be an excellent tracer of disk-halo cycling, its kinematics may shed significant light on the nature of such flows.

Limited studies of EDIG kinematics have been performed to date. \citet{frat04} use spectra from slits along the major and minor axes of NGC 2403 to detect a rotational lag in the extraplanar ionized gas component. Spectra have also been obtained along slits perpendicular to the major axes of NGC 891 \citep[one slit;][]{rand97} and NGC 5775 \citep[three slits;][]{rand00,tull00b}. In both galaxies, the mean velocities are seen to approach the systemic velocity as height above the midplane ($z$) increases. Models representing two distinct physical regimes have been generated to try to understand the apparent drop-off in rotation speed with $z$ in NGC 891 and NGC 5775. Purely hydrodynamical models are described and compared to existing data by \citet{b00} and \citet{barn04}; a purely ballistic model is presented by \citet[][hereafter CBR]{cbr}.

The previous studies of EDIG kinematics were limited by poor velocity resolution and the one-dimensional spatial coverage of long-slit spectroscopy. The mean velocities used are affected by both the rotation speed and the distribution of emission along the line of sight. A full analysis of how gaseous halos rotate requires that these effects be separated as has been done with H$\,$I observations of NGC 891 \citep[e.g.,][]{frat04c}, and this can only be achieved with high spectral resolution and two-dimensional spatial coverage. The work described in this paper represents an extension of the previous optical studies, in that high spectral resolution velocity information is obtained and analyzed for the full two-dimensional extent of NGC 5775.

NGC 5775 is classified as an SBc galaxy. It is undergoing an interaction with its neighbor, NGC 5774, and a tidal stream of H$\,$I connects the two galaxies \citep[see][]{irwin94}. \citet{coll00} cite a high far-infrared luminosity determined from the \emph{IRAS} satellite, $L_{\mathrm{FIR}}=7.9\times 10^{43}$ erg s$^{-1}$, and a ``far-infrared surface brightness'' of $L_{\mathrm{FIR}}/D_{25}^2=8.4\times 10^{40}$ erg s$^{-1}$ kpc$^{-2}$ (where $D_{25}$ is the optical isophotal diameter at 25$^{\mathrm{th}}$ magnitude), which is typical of mild starbursts \citep{rd03a}. The EDIG layer is bright, with a large scale height and filamentary structures \citep{coll00}. A summary of galaxy parameters for NGC 5775 is presented in Table \ref{n5775pars}.

\begin{deluxetable}{ll}
\tablecaption{Galaxy Parameters for NGC 5775}
\tablehead{\colhead{Parameter} & \colhead{Value}}
\tablecolumns{2}
\startdata
RA (J2000.0) & $14^{\mathrm{h}}53^{\mathrm{m}}57\fs 57$ \\
Decl. (J2000.0) & $03^{\mathrm{d}}32^{\mathrm{m}}40\fs 1$ \\
Adopted Distance\tablenotemark{a} & 24.8 Mpc \\
Inclination & 86$\degr$ \\
Position Angle & 145.7$\degr$ \\
Systemic Velocity & 1681.1 km s$^{-1}$ \\
\enddata
\tablenotetext{a}{Assuming $H_0=75$ km s$^{-1}$ Mpc$^{-1}$}
\tablecomments{All values are from \citet{irwin94}.}
\label{n5775pars}
\end{deluxetable}

This paper is organized as follows. We describe the observations and the data reduction steps in \S \ref{observations}. In \S \ref{modeling}, we present an analysis of disk and halo rotation. We compare the data with the ballistic model of CBR in \S \ref{ballistic}, and conclude the paper in \S \ref{conclusion}.

\section{Observations and Data Reduction}

\label{observations}

Data were obtained during the nights of 2001 April 11--13 at the Anglo-Australian Telescope (AAT). The TAURUS-II Fabry-Perot interferometer, which is placed at the Cassegrain focus (f/8) of the AAT, was used in conjunction with the MIT/LL 2k$\times$4k CCD. Design, theory, and data reduction techniques of Fabry-Perot interferometers are well described in the literature; see e.g. \citet{bt89,jones02,gordon00}. An order blocking filter (6601/15) isolated H$\alpha$ emission at order 379 over a 9 arcminute field of view. At this order, the spectral resolution is quite high (FWHM $\simeq$ 0.5\AA\ $=22.9$ km s$^{-1}$), but the wavelength satisfying the etalon interference condition varies radially across a given image (each of which corresponds to a given etalon spacing, $h$). Over the course of the observing run, the etalon spacing was changed to sample the full free spectral range (FSR = 17.4\AA ) at each CCD pixel. It should be noted, however, that telescope pointing variations resulted in the FSR being incompletely sampled at some spatial locations, resulting in blank pixels later in the reduction process. The exposure time at each of the 72 etalon spacings was 12 minutes. To avoid confusion from ghost images, the galaxy was placed away from the optical center (a faint ghost image can be seen in the upper-left corner of Fig. \ref{skysub}a).

Bias frames, taken during the observing run, were averaged and subtracted from each image. Flat fields were averaged and applied to the images. A ``white light'' cube was also obtained. This reduction step is used to remove any wavelength-dependent flat field structure. However, the white light cube was found everywhere to vary by 2 percent or less. A test application of the white light cube yielded little difference; therefore, this correction was deemed unnecessary. The individual images were arranged in order of etalon spacing and stacked into a data cube. To increase signal-to-noise, the images were spatially binned 2$\times$2 in software. Cosmic ray removal was performed by hand using the IRAF\footnote{IRAF is distributed by the National Optical Astronomy Observatories, which are operated by the Association of Universities for Research in Astronomy, Inc., under cooperative agreement with the National Science Foundation.} task CREDIT. 

The resulting cube contains planes of constant etalon spacing rather than wavelength. The surfaces of constant wavelength are paraboloids, centered on the optical axis and with a curvature constant $K_{\lambda}$ which is dependent on instrumental parameters \citep[see][and our equation \ref{instparms}]{bt89}. In addition, the position of the optical axis on the detector is not known a priori; indeed, the position was found to vary over the course of the observing run, distorting the shape of the constant-wavelength paraboloids. Moreover, telescope pointing variations result in shifts of object locations with respect to both the detector and optical centers. The process of aligning the optical axis and transforming the surfaces of constant wavelength to planes is called the phase correction. Our procedure for subtracting night-sky line emission and obtaining the wavelength solution (representing the bulk of the phase correction) is outlined in the Appendix.

Relative intensity variations between the sky-subtracted images were measured and corrected for using field stars. The IRAF task GAUSS was used to convolve each image to a common beam corresponding to the worst seeing conditions from the observing run ($3.55\arcsec$). Because we are primarily looking for diffuse emission, the poor seeing is not harmful; rather, the faint diffuse structure is brought out by the convolution.

Astrometric solutions for each image were calculated with the IRAF task CCMAP, using six relatively bright stars included in the HST Guide Star Catalog\footnote{The Guide Star Catalog was produced at the Space Telescope Science Institute under U.S. Government grant. These data are based on photographic data obtained using the Oschin Schmidt Telescope on Palomar Mountain and the UK Schmidt Telescope.}. With these astrometric solutions, and the wavelength solution described in the Appendix, the phase correction was completed by sorting each intensity value into the correct pixel in the final data cube. In cases where multiple intensity values were assigned to a single pixel, the average value was used. The pixels in the final data cube are $2\arcsec$ square, and the channel width is 11.428 km s$^{-1}$.

With a FSR of only 17.4\AA , corresponding to $\sim$800 km s$^{-1}$ at H$\alpha$, the rotational velocity of the galaxy \citep[198 km s$^{-1}$;][]{irwin94} together with the broad emission profiles ($\sigma_{\mathrm{gas}}\approx 32.5 - 42.5$ km s$^{-1}$; see \S \ref{diskrotcurs}) causes real H$\alpha$ emission to appear in nearly all planes of the data cube. Thus, there are few continuum channels in our data set and the continuum must be removed locally. To remove the continuum emission, a median-filtering algorithm was implemented, using intensity values along the spectral axis at a particular spatial location and those along the four nearest-neighbor spectra. Taking the intensity values from all five spectra together, the median and standard deviation were calculated. Values differing from the median value by more than two standard deviations were excluded (thus eliminating the part of the spectrum containing H$\alpha$ emission), and new statistics were calculated. This procedure was repeated until no more statistical outliers were found, or for a maximum of five iterations. The resulting median was taken as a measure of the continuum level at that location, and was subtracted from the spectrum.

To enhance faint emission far from the galaxy midplane, the data cube was further smoothed to an $8\arcsec$ beam. Blank (unsampled) pixels in the final cube, caused by variations in telescope pointing, were replaced before smoothing by interpolating over adjacent pixels. In the region of the data cube containing emission from NGC 5775, approximately 3 per cent of the pixels were initially blank.

To obtain a rough intensity calibration, a major axis cut through the moment-0 map shown in Fig \ref{mom}a was compared to a major axis cut through the H$\alpha$ image from \citet{coll00}, smoothed to the same resolution as in the moment-0 map. The noise in the channel maps was thus measured to be $1.38 \times 10^{-19}$ ergs cm$^{-2}$ s$^{-1}$ arcsec$^{-2}$ channel$^{-1}$. Assuming a gas temperature T=$10^4$K, this corresponds to an emission measure (EM) per channel of 0.0688 pc cm$^{-6}$ channel$^{-1}$. 

\begin{figure}
\plottwo{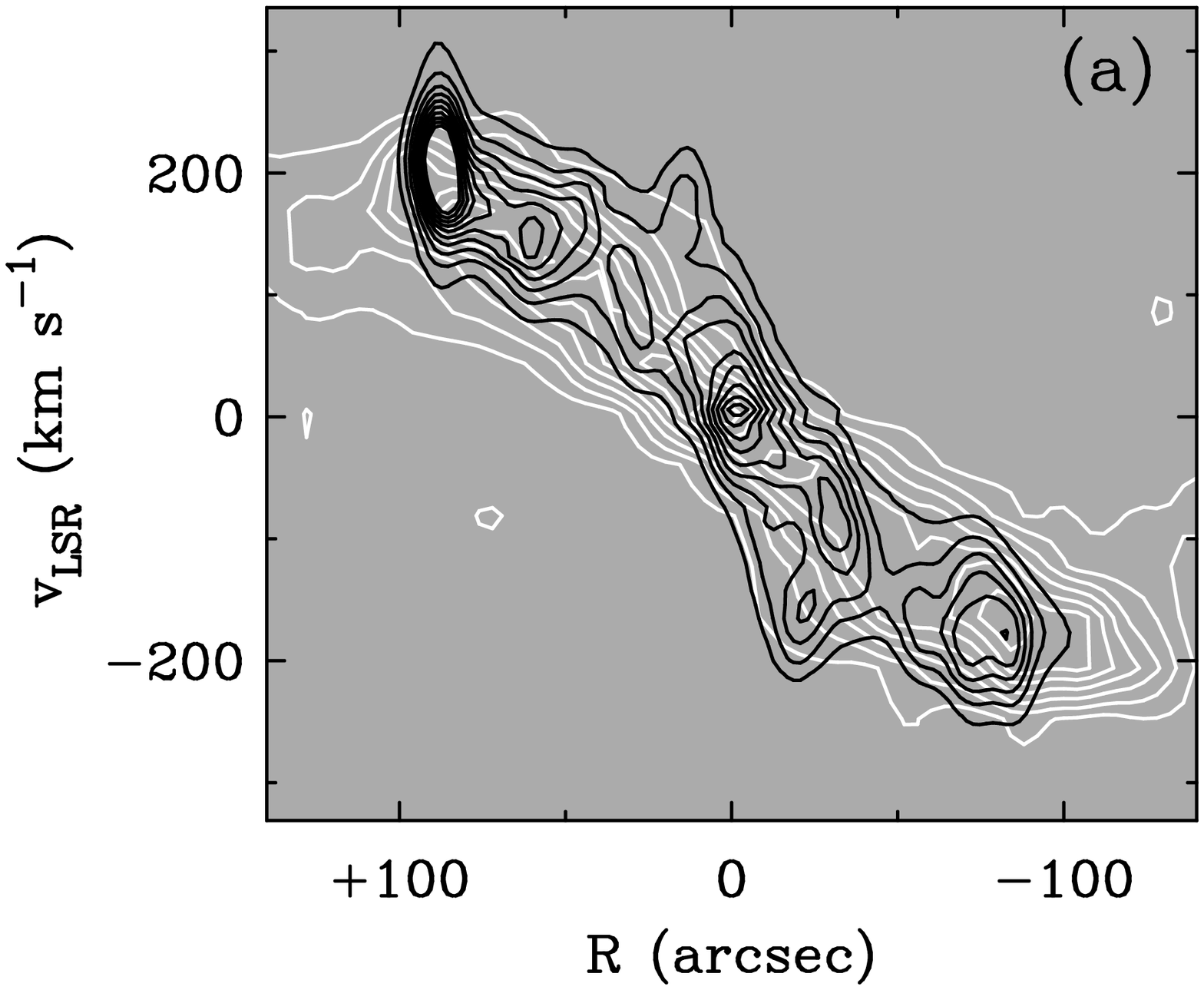}{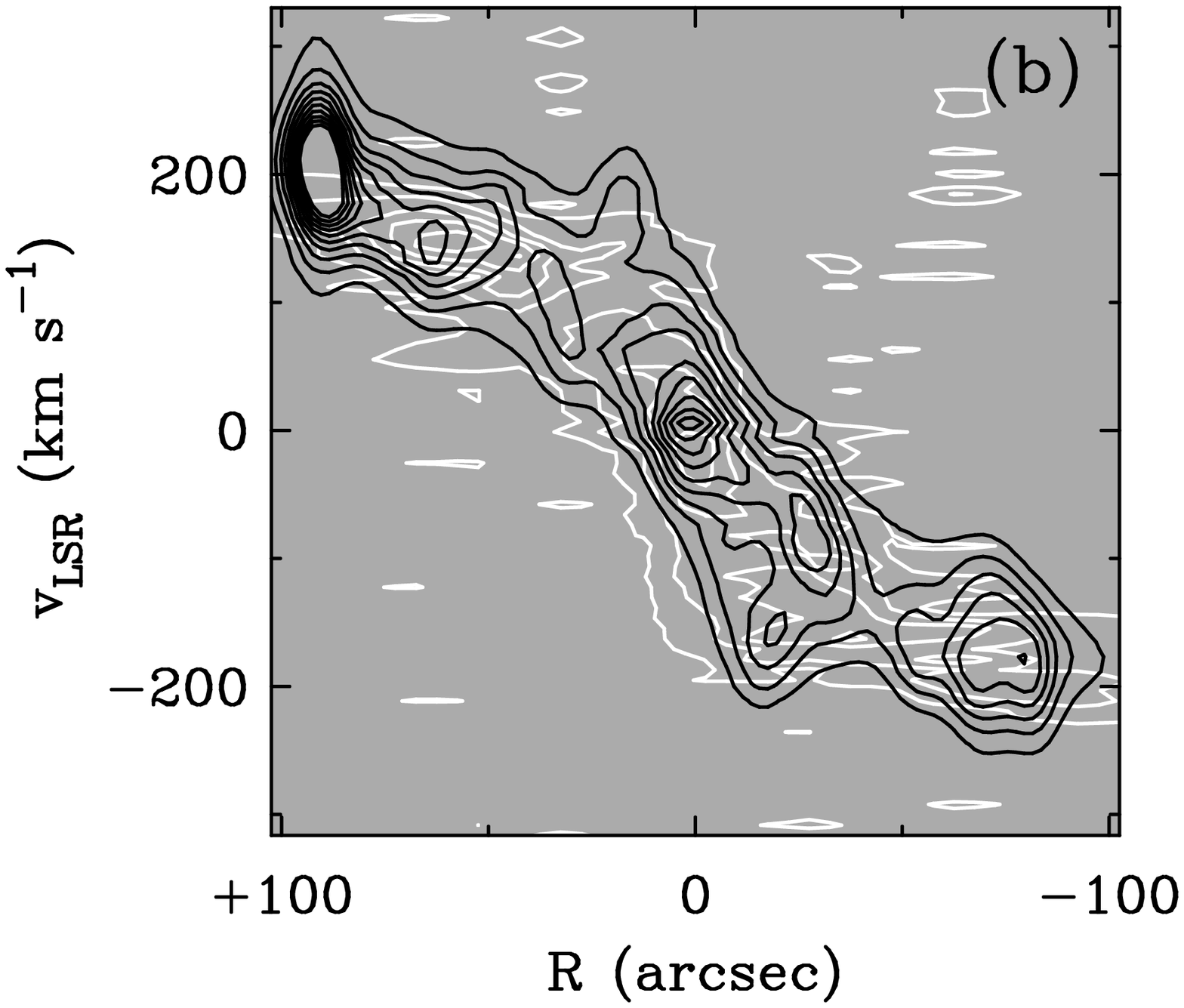}
\caption{(a) Overlay of H$\,$I (\emph{white contours}) and H$\alpha$ (\emph{black contours}) major axis PV diagrams. H$\,$I contours run from 3.36 to 43.7 K in increments of 6.72 K. H$\alpha$ contours run from $8.28\times10^{-18}$ to $7.04\times10^{-17}$ erg cm$^{-2}$ s$^{-1}$ arcsec$^{-2}$ channel$^{-1}$, in increments of $6.90\times10^{-18}$ erg cm$^{-2}$ s$^{-1}$ arcsec$^{-2}$ channel$^{-1}$. (b) Overlay of CO 2--1 (\emph{white contours}) and H$\alpha$ (\emph{black contours}) major axis PV diagrams. CO contours run from 0.03 to 0.33 K in increments of 0.05 K. H$\alpha$ contours are the same as in (a). Positive values of the major axis distance $R$ correspond to the southeast side of the disk. The 9 km s$^{-1}$ offset described in the text has already been applied to the H$\alpha$ velocity scale.}
\label{hico}
\end{figure}

\begin{figure}
\plottwo{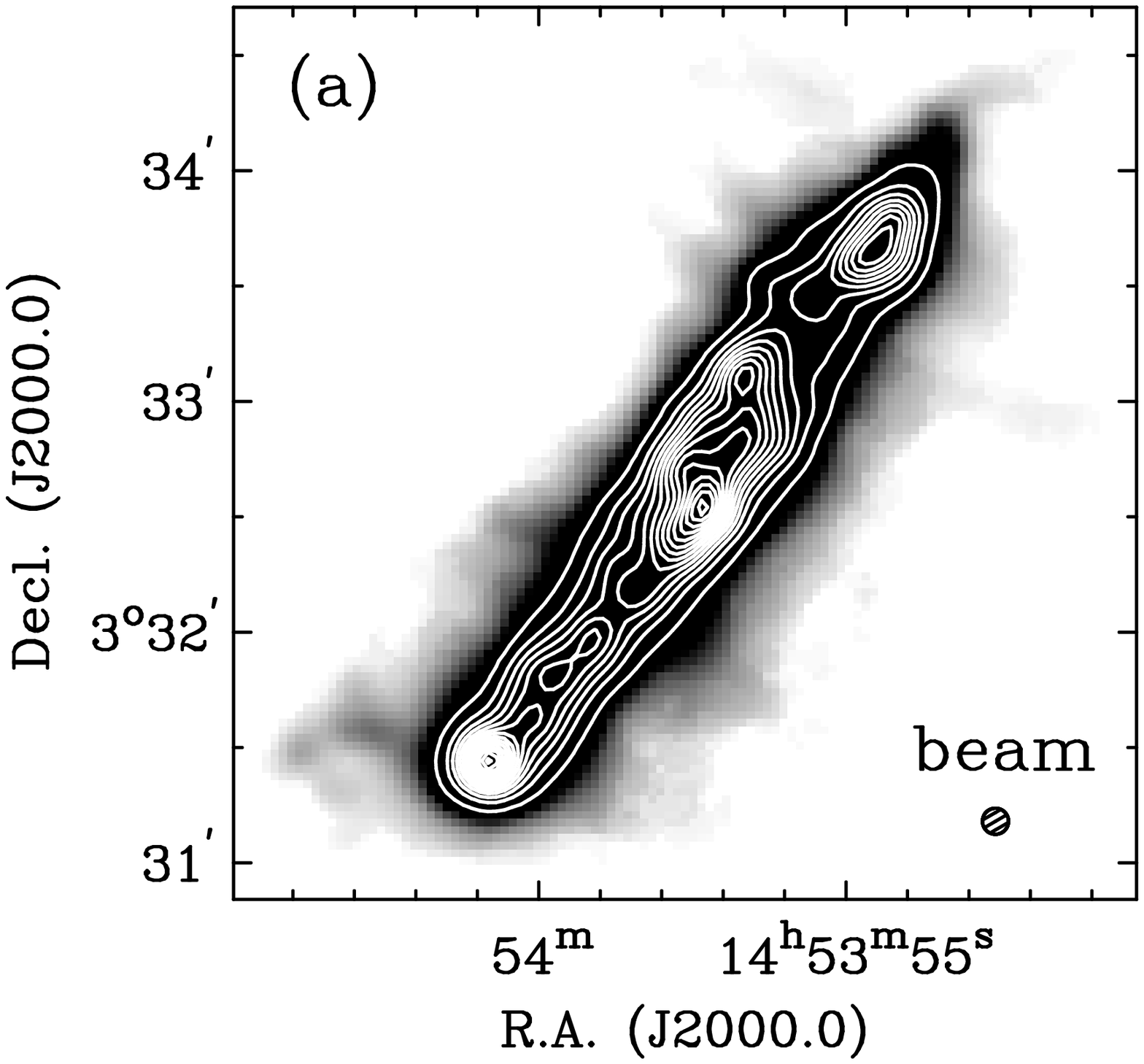}{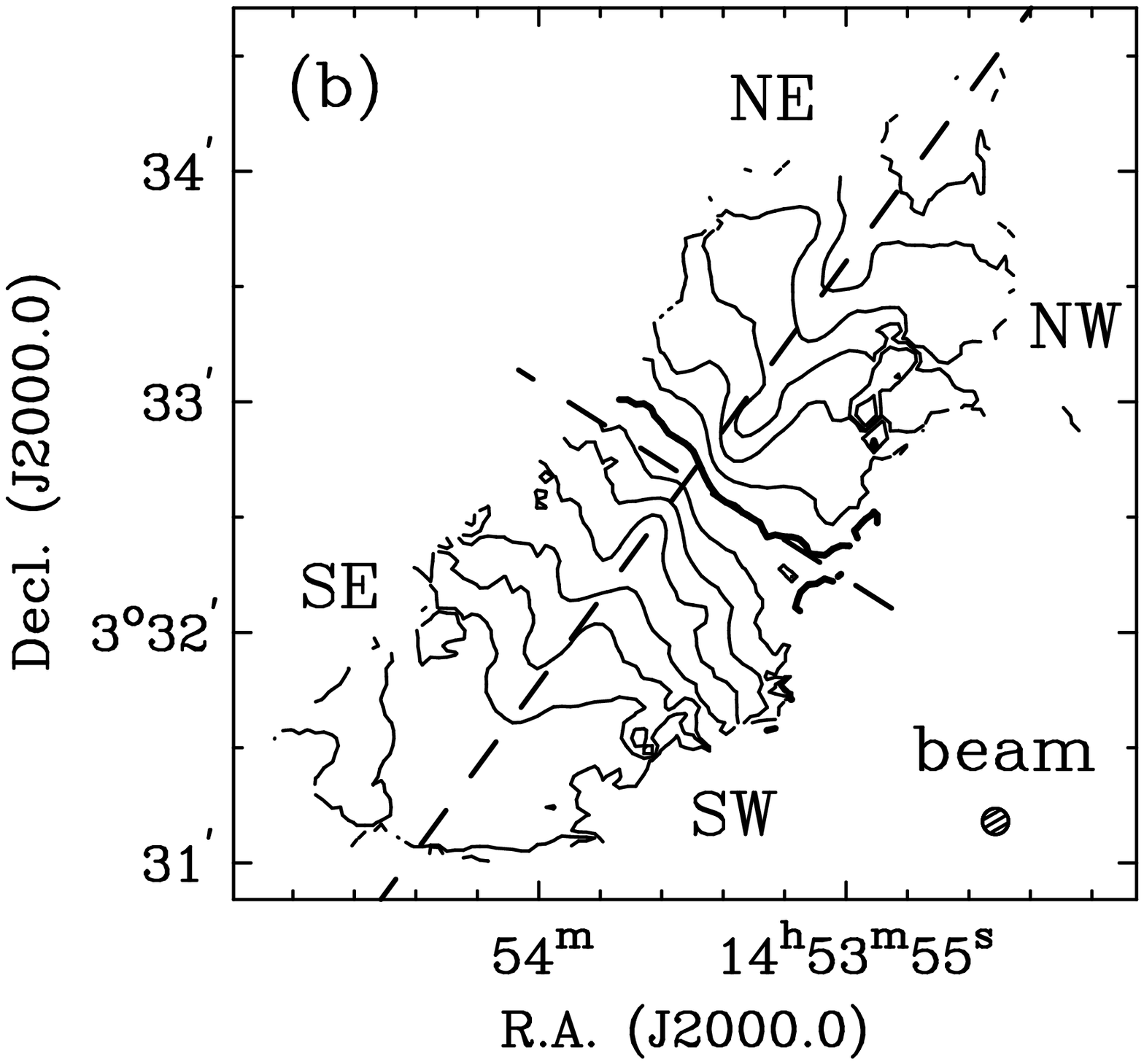}
\caption{(a) Moment-0 map of the Fabry-Perot data cube. Contours illustrate the appearance of the brighter structures in the disk. (b) Moment-1 map of the Fabry-Perot data cube. Velocities are relative to the systemic velocity of 1681.1 km s$^{-1}$. Contours run from -210 to 210 km s$^{-1}$ in increments of 30 km s$^{-1}$ (the SE and SW quadrants constitute the receding side). The systemic velocity contour is darkened. The morphological major and minor axes, determined using the center and position angle listed in Table \ref{n5775pars}, are indicated by dashed lines.}
\label{mom}
\end{figure}

Because the wavelength solution was based on night-sky emission lines rather than observations of a standard wavelength calibration lamp (see the Appendix), the velocity scale was verified by comparing with existing data. H$\,$I data \citep[see][]{irwin94} and CO 2--1 data \citep[see][]{lee01} were kindly provided by J. Irwin  and S.-W. Lee for this purpose. The beam size of the H$\,$I data is $13.6\arcsec \times 13.4\arcsec$ at position angle $-33.7\degr$, and the channel width is 41.67 km s$^{-1}$. The beam size of the CO 2--1 data is $21\arcsec$, and the channel width is 8.08 km s$^{-1}$. Overlays of major axis position-velocity (PV) diagrams were generated to compare both data sets to the H$\alpha$ data. The H$\,$I PV diagram differs substantially from the H$\alpha$ PV diagram in that it shows a slower rise in velocity at low $R$ (see Fig. \ref{hico}a). This is due to the H$\,$I being less centrally concentrated than the ionized gas, as found through modeling of the H$\,$I data cube by Irwin (1994). The H$\,$I diagram is therefore not optimal as a check on the velocity scale.  On the other hand, the CO emission is expected to follow the ionized gas distribution more closely \citep[e.g.,][]{ry99,wb02}, except where extinction may affect the H$\alpha$ profiles. Indeed, the shapes of the PV diagrams are more similar (see Fig. \ref{hico}b). Based on the comparison between the CO and H$\alpha$ PV diagrams (after converting the H$\alpha$ velocities to the LSR frame), a constant 9 km s$^{-1}$ offset was added to the velocity axis of the H$\alpha$ data cube. This correction is smaller than the channel width of the data cube. In all presentations of kinematic data in this paper, velocities are relative to the systemic velocity of NGC 5775 (cf. Table \ref{n5775pars}).


Moment maps were generated with the Groningen Image Processing System (GIPSY) task MOMENTS by requiring that emission appears above the 3$\sigma$ level in at least three velocity channels. The moment-0 (total intensity) map is displayed in Fig. \ref{mom}a, and the moment-1 (mean velocity) map is displayed in Fig. \ref{mom}b.


\section{Analysis and Modeling}

\label{modeling}

The primary goal of this work is to study the kinematics of the ionized halo of NGC 5775. In this Section, we first present the velocity field and point out some interesting features and trends. Next, major axis rotation curves are obtained for both the ionized and molecular gas components from PV diagrams. Our analysis of the kinematic structure is then extended by modeling the halo component to search for a change in the rotation curve with height above the midplane.

\subsection{The velocity field}

\label{velfieldsection}

Some insight into the global kinematic characteristics of the ionized component of NGC 5775 can be obtained by examining the velocity field. Figure \ref{mom}b shows the moment-1 map calculated from the data cube. The following features are observed:
\begin{itemize}
\item In all four galaxy quadrants, mean velocities are seen to decrease with increasing distance from the major axis (up to $\sim 10\arcsec$). This initial decrease is attributed to viewing a rotating disk in projection, and is not indicative of a halo lag. To determine the location of the edge of the projected disk, we assume that the disk is axially symmetric, with a radius equal to the semi-major axis length of $\approx 100\arcsec$. Beyond this distance, the emission is observed to fall sharply, and we therefore assume that the disk has a sharp radial cutoff.
\item In the southwest and northeast quadrants (labeled SW and NE in Fig. \ref{mom}b), for most contours the mean velocities continue to decrease beyond the extent of the projected disk. The EDIG in these regions is dominated by filamentary structures \citep[cf.][]{coll00}.
\item In the southeast quadrant (labeled SE in Fig. \ref{mom}b), the velocity contours show an increase in mean velocity above the edge of the projected disk. The contours then appear to run roughly parallel to the minor axis. In the northwest quadrant (labeled NW in Fig. \ref{mom}b), the behavior of the mean velocities is confused by the presence of a closed contour structure, but it appears that some of the contours show a continued drop in mean velocity above the edge of the projected disk, while other contours run parallel to the minor axis.
\item The asymmetry of the velocity contours along the major axis, particularly in the southern (receding) side, is likely caused by significant dust extinction in the disk. The dust lane, which runs parallel to, but offset to the northeast from the major axis, preferentially obscures emission from gas more distant from the observer along those lines of sight. Because the velocity projections along the line of sight are minimized at the near edge of the disk, we observe lower mean velocities to the northeast of the major axis than to the southwest, where extinction is not as extreme. In other words, extinction in the dust lane results in little or no emission reaching us from near the line of nodes.
\item The kinematic and morphological minor axes are not parallel in the projected disk [also noted in H$\,$I by \citet{irwin94}]. We assume that the emission in the projected disk locations is in fact dominated by disk emission. The sense of the offset, together with the inclination angle of the galaxy and the assumption that the near side is tilted to the northeast with respect to the morphological major axis (based on the appearance of the dust lane), is consistent with a radial inflow of gas in the disk, or may indicate the presence of a bar.  If the former, there is no clear indication that it extends beyond the projected disk into the halo.  NGC 5775 is classified as barred in the Third Reference Catalogue of Bright Galaxies \citep[RC3;][]{rc3}, although an examination of the Two Micron All Sky Survey \citep[2MASS;][]{klein94} $K$-band survey image does not suggest bar signatures such as a box- or peanut-shaped bulge.
\end{itemize}

Though the moment-1 map is useful for viewing overall trends, it cannot be used to derive rotation speeds for a nearly edge-on galaxy. Intensity-weighted velocities are affected by the rotation curve, varying projections of the rotation-velocity vector, and the gas distribution along any line of sight and thus are not indicative of rotation speed. A full consideration of the line profile shapes and three-dimensional density distribution (as well as an assessment of extinction) is necessary to derive robust rotation curves in the case of high inclination. 

\subsection{Disk rotation curves}

\label{diskrotcurs}

Here, we wish to construct disk rotation curves from major axis PV diagrams generated from both the H$\alpha$ and CO 2--1 data sets. Several methods have been suggested for recovering the rotation curve of a galaxy. In particular, two useful methods in the study of edge-on spirals are the ``envelope tracing'' method \citep{sancisi79,sofue01} and the ``iteration method'' \citep{takamiya02}. The envelope tracing method calculates the rotation curve using the high-velocity edge of a PV diagram. The iteration method automates the procedure of generating a model galaxy (with specified radial density profile and rotation curve) that best matches the major axis PV diagram of the observed galaxy. The benefit of using the iteration method is a more accurate recovery of the rotation curve at small radii, which, because of beam smearing and rapidly changing densities and velocities, is typically underestimated by the envelope tracing method. However, our implementation of the iteration method employs the envelope tracing method as part of its fitting procedure; hence, we describe here the specifics of the algorithms used in both methods, which we have written as MATLAB scripts. For reasons discussed in \S \ref{itermethod}, the iteration method was unable to accurately determine the major axis rotation curve for NGC 5775, but the results of the method were used as a starting point for a visual determination.

Throughout this Section, the disk is assumed to consist of a series of concentric, axisymmetric rings, each of which is allowed to have a different gas density, rotational velocity, and velocity dispersion. We do not allow the disk to be warped, under the assumption that the H$\alpha$ emission traces the star forming disk and not the outer parts. An examination of the channel maps did not reveal the signature of a warp along the line of sight \citep[see, e.g.,][]{ssv97}, and the moment-0 map does not show evidence for a warp across the line of sight. Here, we briefly describe the specifics of the algorithms used.

\subsubsection{The envelope tracing method}

\label{envtrace}

After the description of \citet{sofue01}, we use the following algorithm. A PV diagram is generated, and the velocity dispersion ($\sigma_{\mathrm{gas}}$) is estimated by measuring the width of the line profile at the highest radius. It is assumed that at the highest radius, only one ring of emission is being sampled, so velocity projections from additional rings do not broaden the profile. At each radius, the value of the rotation curve is taken to be the velocity at the edge of the line profile, with a correction for the velocity dispersion, the channel width, and the inclination angle of the galaxy.

Mathematically, the intensity of the envelope on the line profile is defined to be \citep{sofue01}
\begin{equation}
I_{\mathrm{env}} = \sqrt{(\eta I_{\mathrm{max}})^2 + I_{\mathrm{lc}}^2},
\end{equation}
where $I_{\mathrm{max}}$ is the maximum intensity in the line profile, $I_{\mathrm{lc}}$ is the lowest contour value, typically taken to be 3 times the rms noise in the PV diagram, and $\eta$ is a constant, normally taken to be in the range 0.2--0.5. On the side of the line profile farthest from the systemic velocity, the rotational velocity is taken to be \citep{sofue01}
\begin{equation}
v_r = (v_{\mathrm{env}} - v_{\mathrm{sys}})/\sin(i) - \sqrt{\sigma_{\mathrm{inst}}^2 + \sigma_{\mathrm{gas}}^2},
\end{equation}
where $v_{\mathrm{env}}$ is the velocity at which $I=I_{\mathrm{env}}$, $v_{\mathrm{sys}}$ is the systemic velocity, $i$ is the inclination angle, and $\sigma_{\mathrm{inst}}$ is the instrumental velocity resolution.

\subsubsection{The iteration method}

\label{itermethod}

Our implementation of the iteration method makes use of the envelope tracing method and the suite of GIPSY tasks. Most importantly, the task GALMOD is used to generate a model data cube using user-specified radial and vertical density profiles, rotation curve, velocity dispersion, and viewing angle. To create the initial model, we specify: the velocity dispersion, which is estimated in the same way as for the iteration method; the radial density profile, which is estimated using the GIPSY task RADIAL; and the initial guess for the rotation curve, which is estimated by using the envelope tracing method on the observed data set.  The radial profile must be well matched to the data because both the density distribution and the rotation curve affect the shape of the line profiles. The scale of the input radial density profile is fixed such that the signal-to-noise in the model is approximately that measured in the data. To measure the noise in the model, the same inputs are used with two different random number seeds, and the standard deviation of the difference between the two runs of the model is calculated. 

Once the initial model has been generated, we follow a procedure based on that detailed by \citet{takamiya02}. The envelope tracing method is used to derive a rotation curve from the initial model. That rotation curve is compared to the one derived from the data, and the differences between the two are added to the input rotation curve to generate a new model. This procedure is repeated until the difference between the rotation curves derived from the data and the most recent model satisfies a convergence criterion or until a maximum number of iterations have been performed. The degree of success of this method is dependent on how well the other galaxy parameters (radial density profile, velocity dispersion, signal-to-noise ratio) are reproduced in the model. If these parameters are poorly specified, the iteration method will not be able to correctly reproduce the observed PV diagram.

\begin{figure}
\plotone{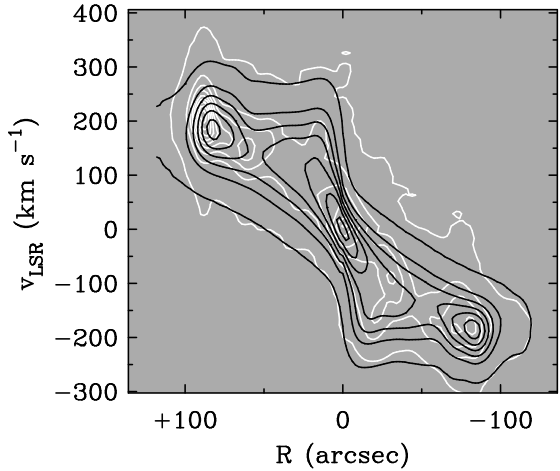}
\caption{Overlay of H$\alpha$ (\emph{white contours}) and best-fit model (\emph{black contours}) major axis PV diagrams. Contour levels for both are 10$\sigma$ to 1010$\sigma$ in increments of 100$\sigma$. Positive values of the major axis distance $R$ correspond to the southeast side of the disk.}
\label{bmma}
\end{figure}

Because RADIAL assumes a circularly symmetric density distribution, it is unable to match the observed major-axis intensity profile, which shows clear signs of clumpiness (see, for example, the bright knot of emission at $R=+90\arcsec$ in Fig. \ref{bmma}). The deviations from axisymmetry are severe enough that the model PV diagrams are too poorly matched to the data to allow the iteration method to converge.
We therefore abandon this method as a means of directly obtaining the rotation curve, and attempt to mitigate some of the effects of the clumpy density distribution. Better results are obtained (though less efficiently) by {\it starting} with the converged radial density profile obtained with RADIAL and a rough rotation curve obtained with the iteration method after a few iterations. These results are used to generate a modeled major axis PV diagram. The observed and modeled PV diagrams are compared by eye, and first the radial density profile is adjusted so that observed and modeled major axis intensity distributions are reasonably well matched except in regions of obvious clumpiness. Then, the rotation curve is modified until a good match is achieved.

\begin{figure}
\plottwo{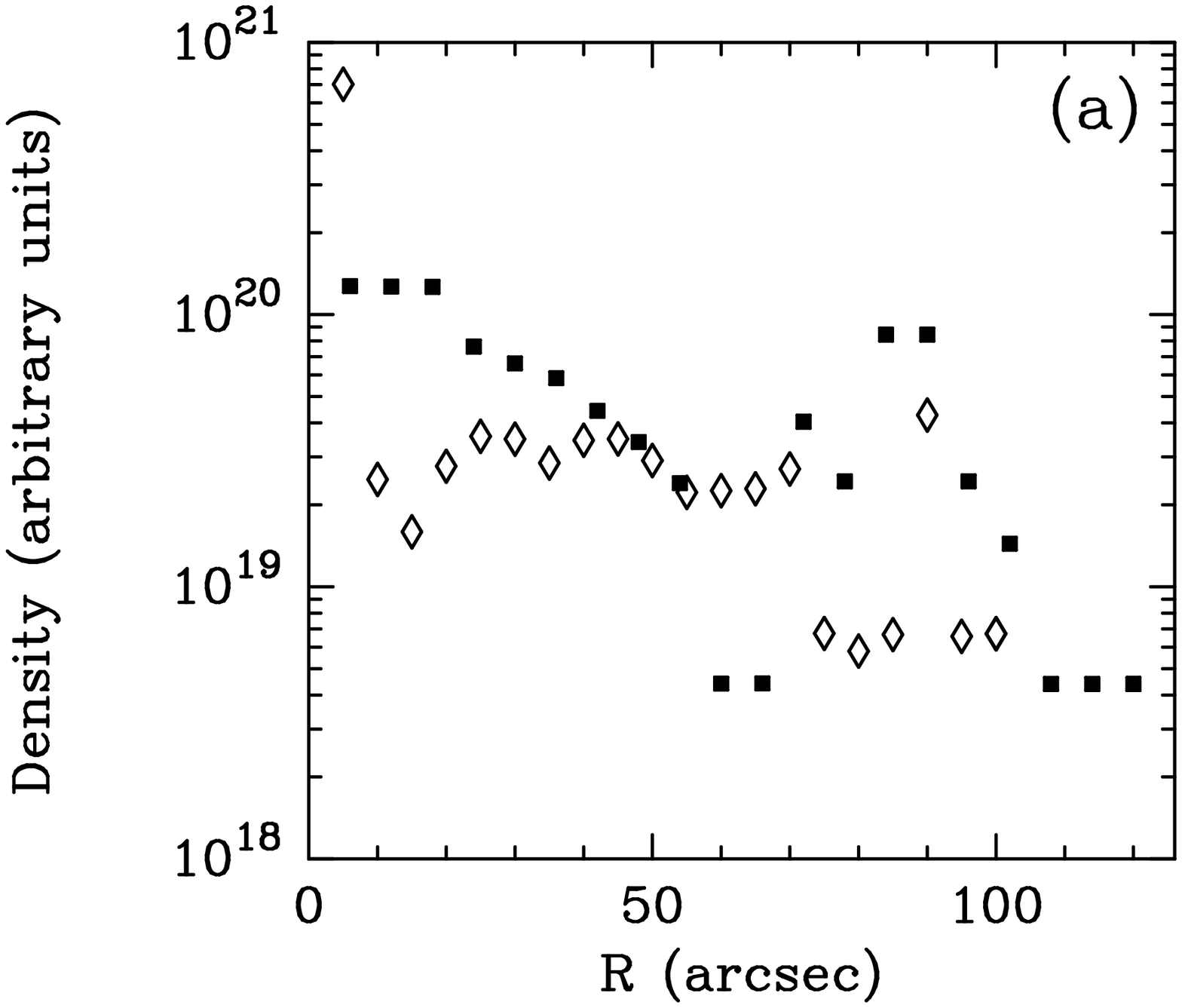}{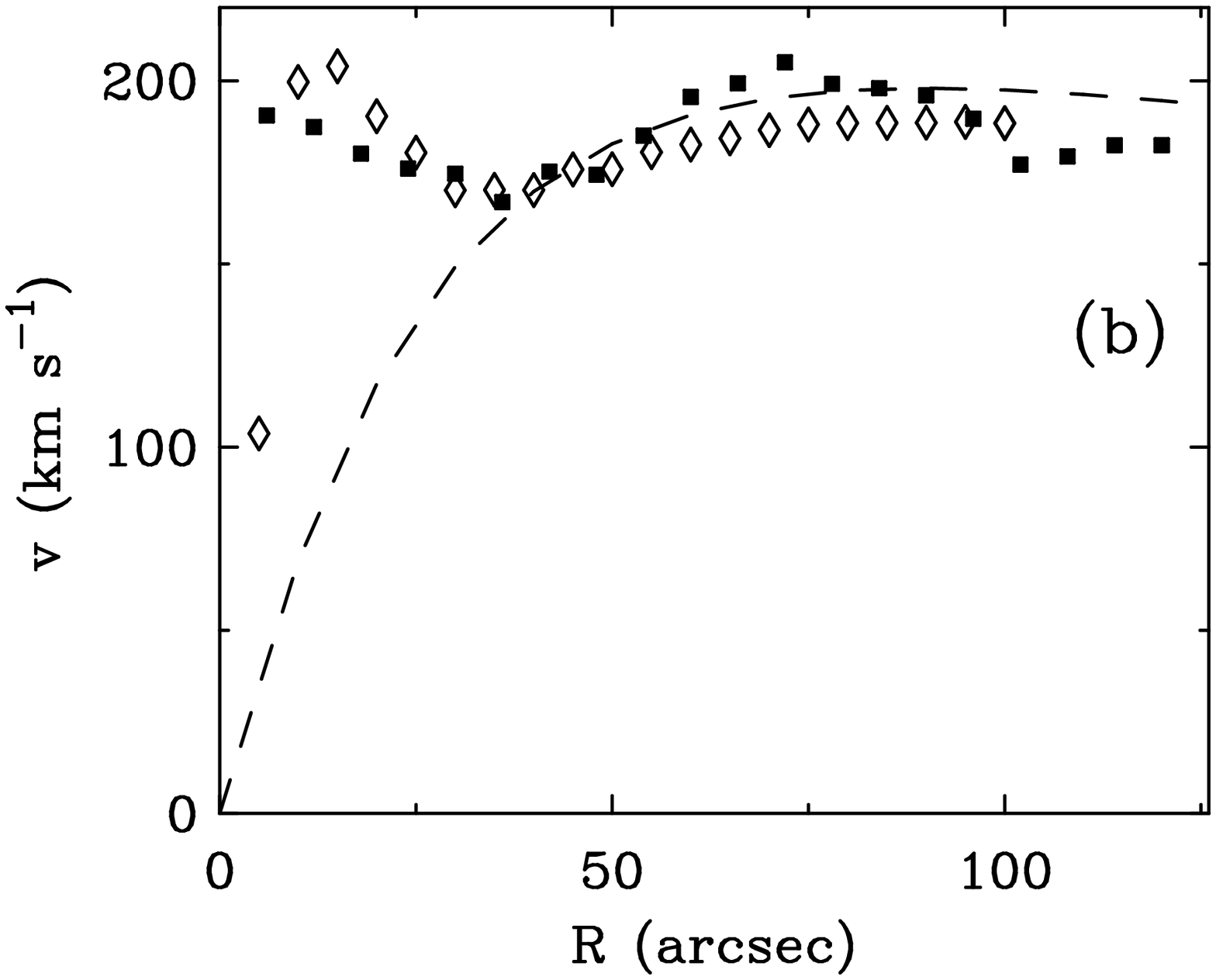}
\caption{(a) Profile of density versus galactocentric radius $R$ for the best-fit models -- H$\alpha$ (\emph{squares}) and CO 2--1 (\emph{diamonds}). (b) Rotation curve for the best-fit models -- H$\alpha$ (\emph{squares}) and CO 2--1 (\emph{diamonds}). The rotation curve of \citet{irwin94} is plotted (\emph{dashed line}) for reference.}
\label{bmrv}
\end{figure}

A major-axis rotation curve and a radial density profile have been obtained in this way from the Fabry-Perot data. Fig. \ref{bmma} shows an overlay of the major axis PV diagrams obtained from the data and the best-fit model. Fig. \ref{bmrv} shows the radial density profile and rotation curve in the best-fit model. To achieve the best match, the velocity dispersion in the model was set to 32.5 km s$^{-1}$ on the approaching side, and 42.5 km s$^{-1}$ on the receding side. The kinematic center and systemic velocity listed in Table \ref{n5775pars} were used, and provided good agreement.


\begin{figure}
\plotone{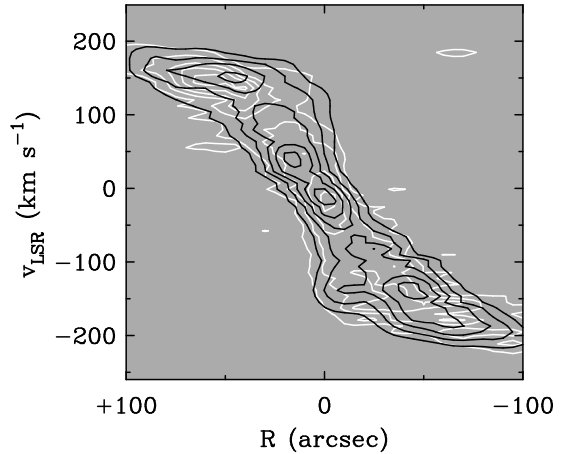}
\caption{Overlay of CO 2--1 (\emph{white contours}) and best-fit model (\emph{black contours}) major axis PV diagrams. Contour levels for both are 3$\sigma$ to 18$\sigma$ in increments of 3$\sigma$. Positive values of the major axis distance $R$ correspond to the southeast side of the disk.}
\label{cobm}
\end{figure}

To check that the radial density profile and rotation curve are reasonable and that our results are not biased by, for example, extinction, we have repeated this procedure for the CO 2--1 data. Fig. \ref{cobm} shows an overlay of the major axis PV diagrams obtained from the data and the best-fit model. Fig. \ref{bmrv} shows the radial density profile and rotation curve in the best-fit model. A velocity dispersion of 15 km s$^{-1}$ was required to obtain the best match. Good agreement was obtained without modifying either the systemic velocity or the kinematic center listed in Table \ref{n5775pars}. In order to match the low signal-to-noise ratio of the CO data, the modeled emission is quite faint; the apparent asymmetry between the approaching and receding sides of the model contours in Fig. \ref{cobm} is due to the noise in the model.

Despite the fact that our H$\alpha$ and CO radial profiles differ, the rotation curves are very similar. The mean difference between the H$\alpha$ and CO rotation curves is 1.5 per cent, the maximum difference is 29.5 per cent (at low $R$, where the CO rotation curve rises more slowly than does the H$\alpha$ rotation curve), and the rms difference is 11.5 km s$^{-1}$. This correspondence implies that our procedure works well. We note that the match between the H$\alpha$ PV diagram and the corresponding model is worse in the inner parts ($R\lesssim 20\arcsec$) than for the CO PV diagram. In the former, the data appear to rise more slowly with a higher velocity dispersion.


\subsection{Halo rotation}

\label{halorotation}

Having recovered a radial density profile and rotation curve for the major axis, we now move on to modeling the halo H$\alpha$ emission. PV diagrams are constructed along cuts parallel to the major axis, at various heights above the midplane $z$ (in this paper, $z$ is positive to the southwest of the major axis). To calculate model PV diagrams, we have made a modification to the GIPSY task GALMOD to allow for a vertical gradient in azimuthal velocity:
\begin{equation}
v(R,z) = \left\{
\begin{array}{ll}
v(R,z=0) - \frac{dv}{dz} [|z|-z_0] & \mathrm{for\ } z > z_0\nonumber\\
v(R,z=0) & \mathrm{for\ } z \leq z_0,
\end{array}
\right.
\label{dvdzeqn}
\end{equation}
where $dv/dz$ is a constant parameter in the model, with units [km s$^{-1}$ arcsec$^{-1}$]. Note that this model fixes the shape of the rotation curve as a function of height, and only changes the amplitude. We include a parameter ($z_0$) that specifies the height at which the rotational lag begins in order to be consistent with \citet{frat04c}, who find that the neutral halo of NGC 891 co-rotates with the disk up to $z=1.3$ kpc, and shows a vertical gradient in azimuthal velocity above that height (although the authors state that it is possible that the observed corotation at this height is an effect of beam smearing). Limited information is available to constrain whether $z_0$ differs from zero in our model. For NGC 5775, a height of 1.3 kpc corresponds to $z=11\arcsec$. However, evidence discussed in this Section indicates that a gradient is already present at $z=10\arcsec$. We therefore set $z_0$ to be the (exponential) scale height of the galaxy model (of order $5\arcsec$; see below), and show later that a choice of $z_0=0\arcsec$ does not change the derived gradient significantly.

Such a vertical gradient in azimuthal velocity is considered in \S \ref{vvgsection}. We also consider the effects of modifying the shape of the halo radial density profile (\S \ref{rdenssection}), the shape of the halo rotation curve (\S \ref{vrotsection}), and the position angle and inclination of the halo (\S \ref{halopa}).

Assuming a circularly symmetric disk with a sharp optical cutoff radius of $100\arcsec$ (cf. Figure \ref{mom0cutfig}d) and an inclination angle of $86\degr$, the edge of the projected disk along the minor axis is located at a distance of $7\arcsec$. \citet{coll00} report that, based on the modeling of \citet{byun94}, extinction effects should be negligible at $z\geq 600$ pc $ = 5\arcsec$. Therefore, at minor axis distances greater than $10\arcsec$, we assume the effects of the projected disk and extinction are negligible. After smoothing the data to an $8\arcsec$ beam, reasonable PV diagrams can be constructed up to a height $|z| \approx 30\arcsec$, but beyond that point there is not enough reliably detected emission. Thus, the range of modeled heights is $10\arcsec \leq |z| \leq 30\arcsec$ (1.2 kpc $\leq |z| \leq$ 3.6 kpc).

\begin{figure}
\begin{center}
\plotone{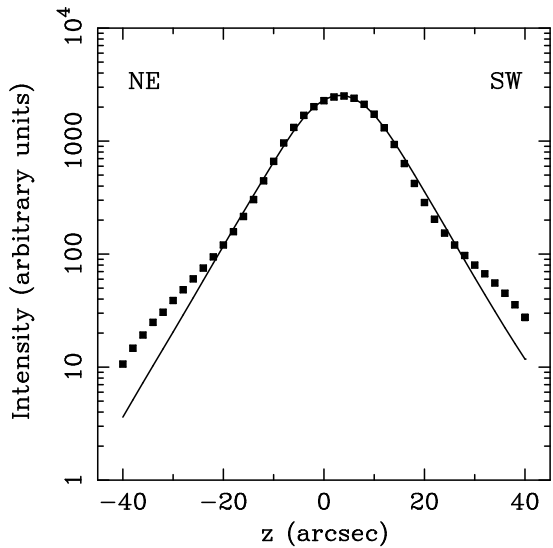}
\end{center}
\caption{Comparison of the observed (\emph{squares}) and modeled (\emph{solid line}) vertical intensity profiles. Both profiles were obtained by averaging the vertical distribution of emission along the length of the disk. The modeled profile is normalized to the observed profile at $z=0$. The exponential scale heights (as described in the text) are $5.8\arcsec$ on the southwest side and $4.5\arcsec$ on the northeast side. At $z=-30\arcsec$, the model is significantly lower than the data; hence the $5.8\arcsec$ scale height is used to generate PV diagrams at that particular height.}
\label{zproffig}
\end{figure}

\begin{figure*}
\begin{center}
\includegraphics[angle=270,scale=0.65]{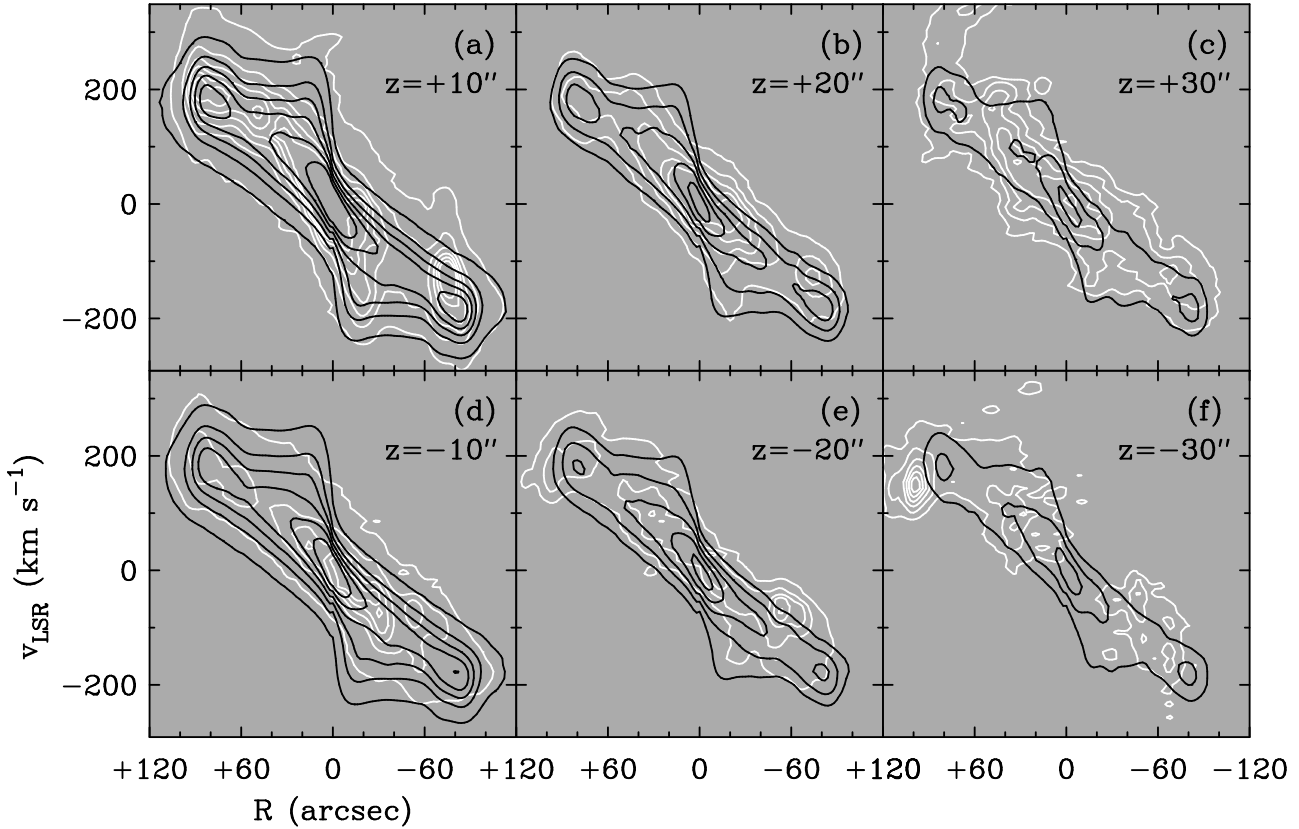}
\end{center}
\caption{PV diagram overlays of H$\alpha$ (\emph{white contours}) and CR model (\emph{black contours}). PV diagrams are displayed at (a) $z=+10\arcsec$, with contour levels 10$\sigma$ to 210$\sigma$ in increments of 40$\sigma$; (b) $z=+20\arcsec$, with contours levels 10$\sigma$ to 50$\sigma$ in increments of 10$\sigma$; (c) $z=+30\arcsec$, with contour levels 3$\sigma$ to 18$\sigma$ in increments of 3$\sigma$; (d) $z=-10\arcsec$, with contour levels 10$\sigma$ to 210$\sigma$ in increments of 40$\sigma$; (e) $z=-20\arcsec$, with contour levels 6$\sigma$ to 36$\sigma$ in increments of 6$\sigma$; (f) $z=-30\arcsec$, with contour levels 3$\sigma$ to 18$\sigma$ in increments of 3$\sigma$.  Positive values of galactocentric radius $R$ and $z$ correspond to the southeast side of the disk and the southwest side of the halo, respectively.}
\label{bmhalo}
\end{figure*}

First, the model described in \S \ref{itermethod} is considered with no vertical gradient in azimuthal velocity. We call this the cylindrical rotation (CR) model. The scale height is adjusted to match the minor axis intensity distributions, but the radial density distribution and rotation curve derived for the major axis are unchanged. The necessary scale heights are $5.8\arcsec$ for the southwest side of the disk ($z > 0$) and $4.5\arcsec$ for the northeast side of the disk (except for the $z=-30\arcsec$ PV diagram, for which the $5.8\arcsec$ scale height was retained). The observed vertical emission profile is compared to the model in Figure \ref{zproffig}. Fig. \ref{bmhalo} displays overlays of the observed PV diagrams and the CR model PV diagrams for slices at heights $z=\pm 10\arcsec, \pm 20\arcsec, \pm 30\arcsec$.


Upon examining these diagrams, some clear features are:
\begin{itemize}
\item A rotation velocity gradient appears to be necessary primarily on the northwest side ($R<0$). We attempt to model this gradient in \S \ref{vvgsection}. In some panels, a shift in systemic velocity seems to be necessary. Such an offset is also included in \S \ref{vvgsection}. The kinematics appear different on the positive-$z$ and negative-$z$ sides of the halo.
\item The ``knee'' in the model PV diagrams, caused by a combination of high gas density and rotational velocity, appears to be unnecessary in some panels, but in most, it appears that this knee moves further from $R=0\arcsec$ with increasing $z$. The shift in this knee may be explained either by a change in the radial density profile (demonstrated in \S \ref{rdenssection}), perhaps due to a radial migration, or by a change in the shape of the rotation curve with height (demonstrated in \S \ref{vrotsection}), as might be expected as the influence of the bulge potential diminishes.  The knee could also be a signature of non-circular disk motions associated with a bar potential, although as mentioned in \ref{velfieldsection}, there is little evidence for a bar in 2MASS images.  
\item In panel \emph{f}, the bright knot of emission at $R\approx +100\arcsec$ and $v\approx 150$ km s$^{-1}$ corresponds to a known \citep[][]{coll00,lee01} H$\alpha$ extension of possibly tidal origin.  This emission is clearly visible in the moment-0 map in Fig. \ref{mom}, extending to the east away from the southeast edge of the disk. That the radial velocity of this complex appears to be significantly lower than that of the surrounding gas either means that the rotational velocity has dropped significantly or that the complex does not lie along the line of nodes.
\end{itemize}

\subsubsection{Azimuthal velocity gradient}

\label{vvgsection}

The slit spectra presented by \citet{rand00} and \citet{tull00b} were examined to provide a first estimate of the magnitude of a possible vertical gradient in azimuthal velocity. Three slits are available, each measuring gas velocities on both sides of the midplane, so that the gradient may be estimated in six regions. Based solely on the mean velocities, the vertical gradient in azimuthal velocity was estimated to be $\sim 1 - 2$ km s$^{-1}$ arcsec$^{-1}$.

\begin{figure*}
\begin{center}
\includegraphics[angle=270,scale=0.65]{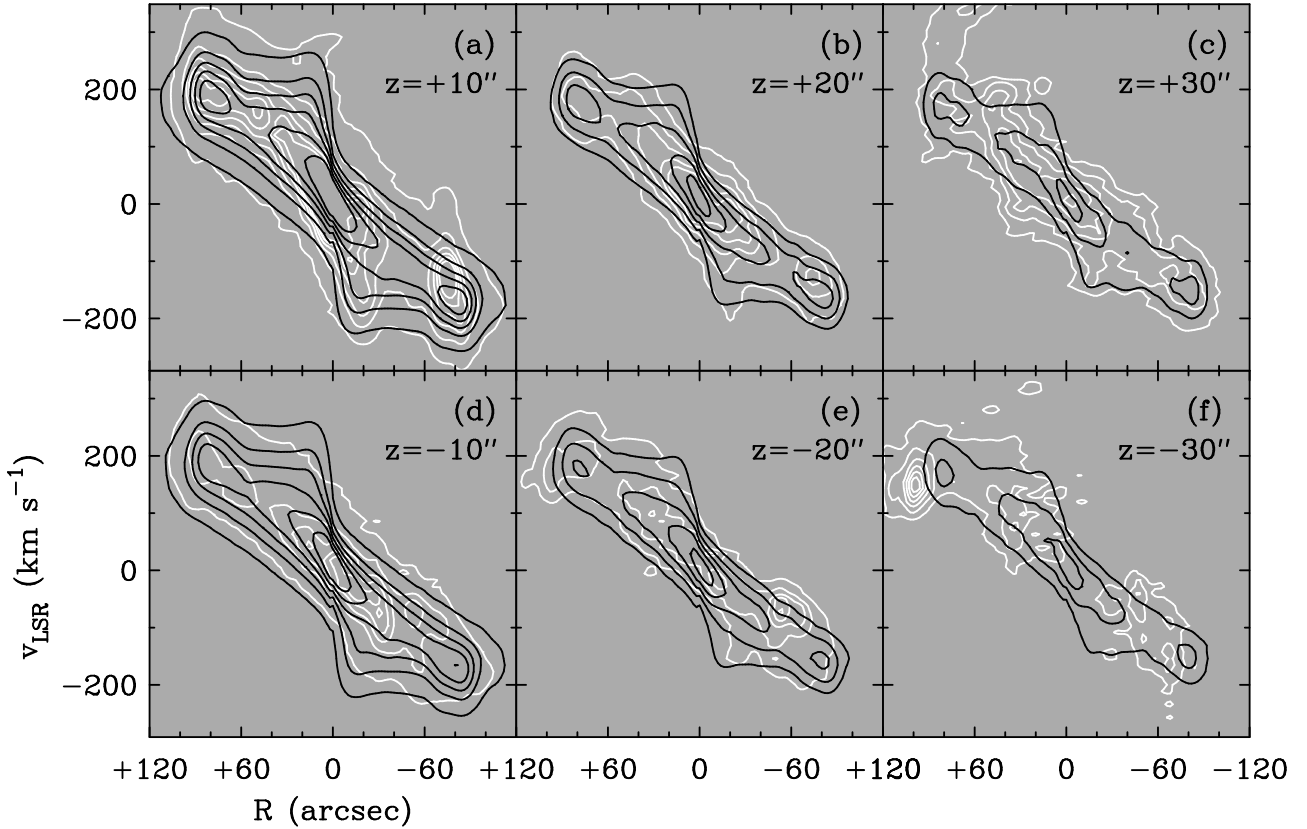}
\end{center}
\caption{PV diagram overlays of H$\alpha$ (\emph{white contours}) and model (\emph{black contours}) where a vertical gradient in azimuthal velocity of 1 km s$^{-1}$ arcsec$^{-1}$ has been added to both sides of the disk, and the systemic velocity has been increased by 10 km s$^{-1}$. Contour levels are as in Fig. \ref{bmhalo}.}
\label{syspv}
\end{figure*}

To test for the existence of a vertical gradient in azimuthal velocity of the form described by equation \ref{dvdzeqn}, we have added a 1 km s$^{-1}$ arcsec$^{-1}$ (or, equivalently, about 8 km s$^{-1}$ kpc$^{-1}$) gradient to the CR model. Figure \ref{syspv} demonstrates the change in shape of the PV diagrams when the gradient is included in the model. The flat part of the approaching side appears to be better matched. However, there is still an apparent problem matching the shape of the PV diagrams for $R\lesssim 60\arcsec$ in the halo (see \S \ref{rdenssection} and \S \ref{vrotsection} for possible explanations). It is not clear that adding the gradient to the receding side has improved the model.

An adjustment of the systemic velocity appeared to be necessary as well in most panels. The CR model was further modified by adding a +10 km s$^{-1}$ offset to the systemic velocity. This offset is already included in Fig. \ref{syspv}. Though the offset is smaller than the channel width of the Fabry-Perot data cube, it seems to improve the match between model and data (for all regions of the halo). Note that this adjustment was not indicated by the major axis PV diagrams; it is only added to improve the agreement between data and model PV diagrams in the halo. We also note that applying a shift in the systemic velocity without retaining the gradient in azimuthal velocity is not sufficient to match the shape of the PV diagrams in the halo.

We have also generated models with different values of the azimuthal velocity gradient. A gradient as low as 0.5 km s$^{-1}$ arcsec$^{-1}$ is found to be too small to match the observed PV diagrams. Gradients with higher amplitude (in particular, $dv/dz=2$ km s$^{-1}$ arcsec$^{-1}$) are able to better match some individual regions (primarily the flat part of the PV diagrams on the approaching side for $z=\pm 20\arcsec$), but provide poor agreement at $z=\pm 30\arcsec$. We conclude that $dv/dz=1$ km s$^{-1}$ arcsec$^{-1}$ provides the best overall agreement with the data, but that higher values of the azimuthal velocity gradient may be appropriate in localized regions of the halo. We note that for a model with this gradient but with $z_0=0\arcsec$, the azimuthal velocities would be approximately 5 km s$^{-1}$ lower at all locations above one scale height, corresponding to a 3 per cent decrease at $z=30\arcsec$ for a flat rotation curve of $v(R,z=0)=200$ km s$^{-1}$.


\subsubsection{Modification of the halo radial density profile}

\label{rdenssection}

\begin{figure}
\begin{center}
\epsscale{0.5}
\plotone{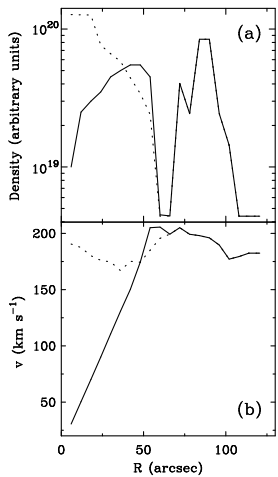}
\end{center}
\caption{(a) Comparison between the profile of density versus galactocentric radius $R$ in the base model (\emph{dotted line}) and the modification presented in \S \ref{rdenssection} (\emph{solid line}). (b) Comparison between the rotation curve in the base model (\emph{dotted line}) and the modification presented in \S \ref{vrotsection} (\emph{solid line}).}
\label{rvmodfig}
\end{figure}

\begin{figure*}
\begin{center}
\includegraphics[angle=270,scale=0.65]{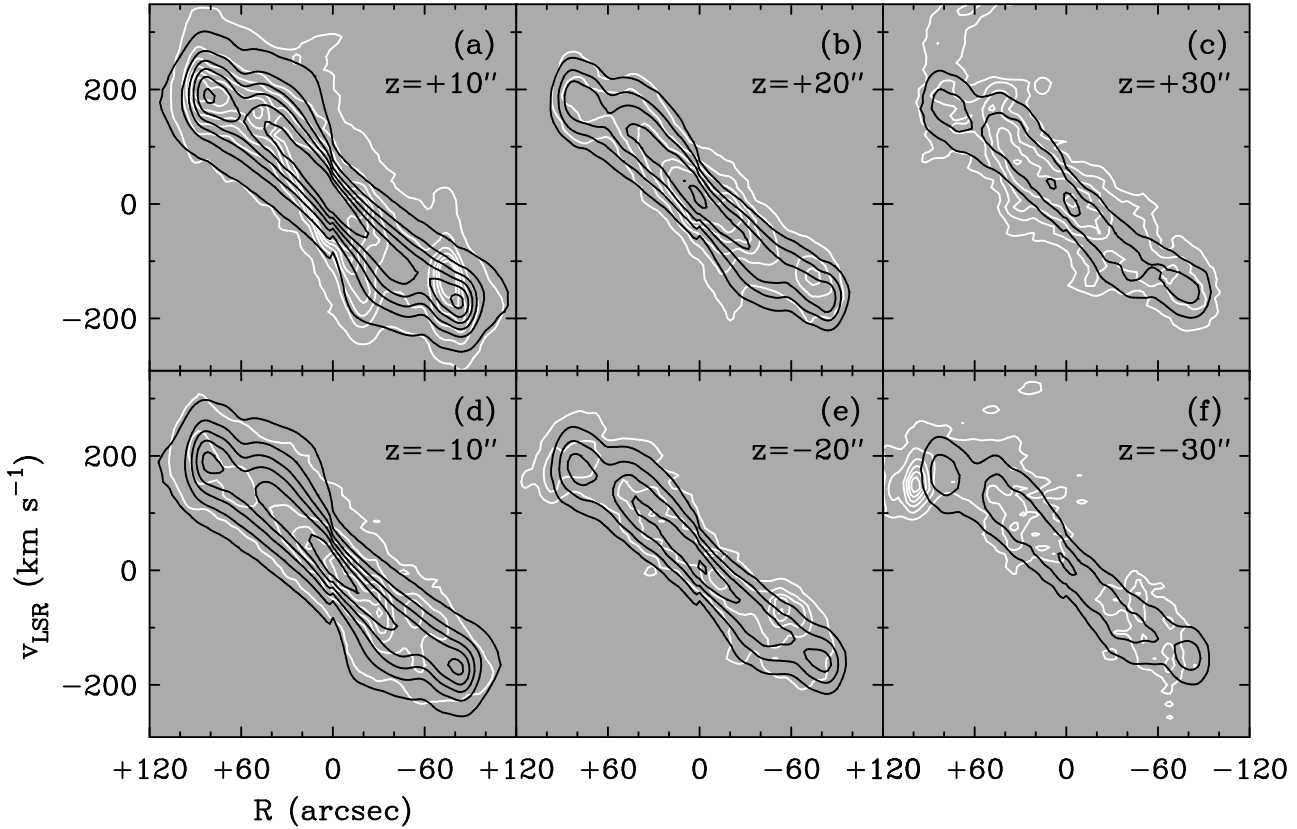}
\end{center}
\caption{PV diagram overlays of H$\alpha$ (\emph{white contours}) and model (\emph{black contours}) where the radial density profile has been varied as shown in Figure \ref{rvmodfig}a. Contour levels are as in Fig. \ref{bmhalo}.}
\label{rdpv}
\end{figure*}

In Figs. \ref{bmhalo} and \ref{syspv}, the ``knee'' in the modeled PV diagrams is too prominent when compared with the data (except at $z=+10\arcsec$, for $R<0\arcsec$). We next make an attempt to match the shape of the PV diagrams in the halo, retaining the features of our best halo model thus far. The shape of the rotation curve is unchanged; only the radial density profile is modified (in the same way for each height $z$), as shown in Figure \ref{rvmodfig}a, by the introduction of a central depression. We have not attempted to recover the actual radial density profile in the halo. Rather, we mean to illustrate that a radial density profile different from that obtained for the major axis better reproduces the changing shape of the PV diagrams with height. The PV diagram overlays are shown in Fig. \ref{rdpv}. We note that the $z=-30\arcsec$ panel suggests a central depression even more severe than we have modeled. Further evidence for a change in the radial gas distribution with height will be discussed in \S \ref{ballistic}: cuts through the moment-0 map (shown in Figure \ref{mom0cutfig}) suggest that such a central hole may well be present in the halo, especially on the negative-$z$ side. Moreover, comparisons with PV diagrams constructed from the ballistic base model (discussed in \S \ref{ballistic}; see Figure \ref{bmpvfig}) suggest that the changing shape of the observed PV diagrams is strongly affected by radial redistribution of matter in the halo.


\subsubsection{Modification of the halo rotation curve}

\label{vrotsection}

\begin{figure*}
\begin{center}
\includegraphics[angle=270,scale=0.65]{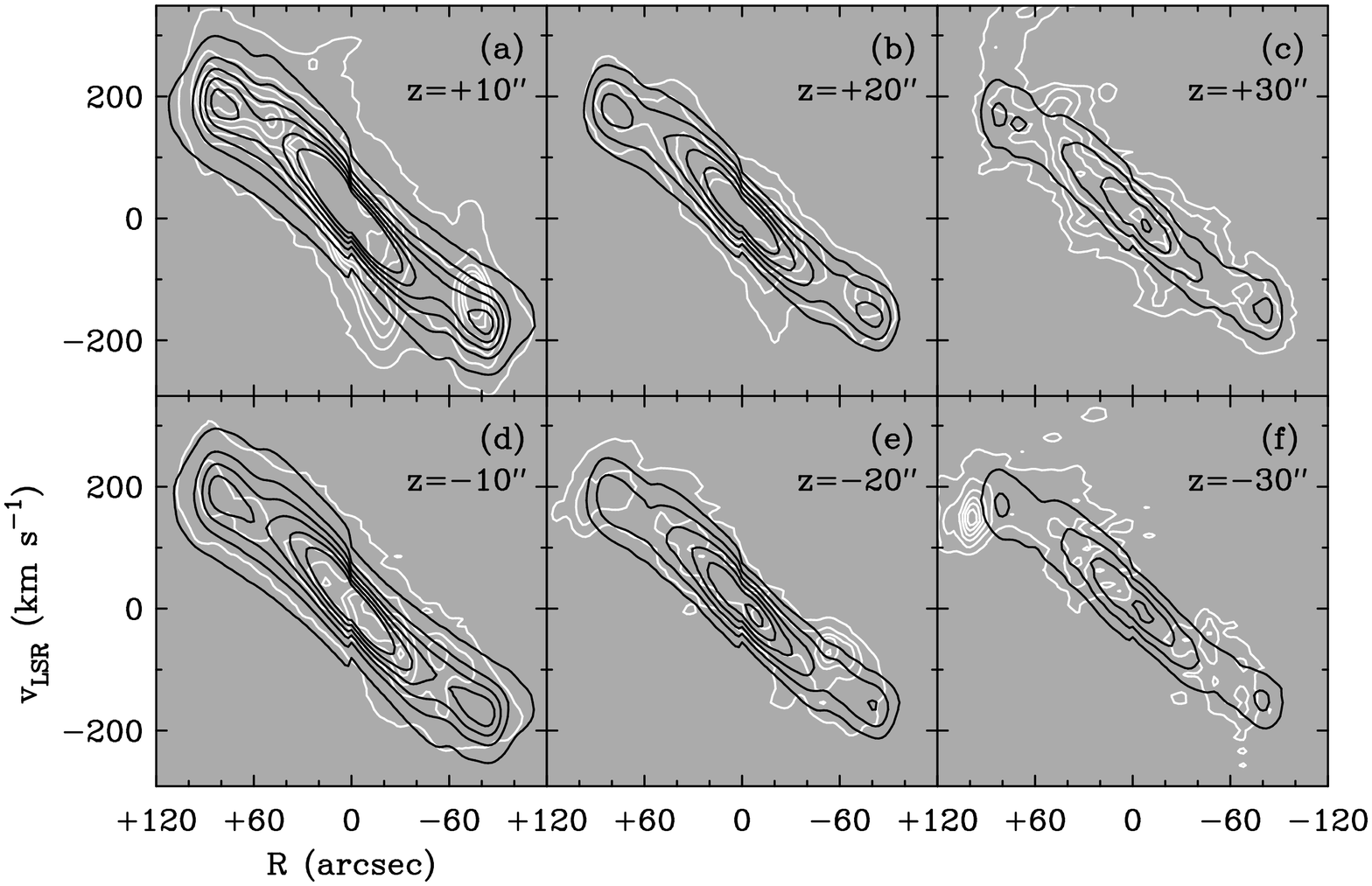}
\end{center}
\caption{PV diagram overlays of H$\alpha$ (\emph{white contours}) and model (\emph{black contours}) where the shape of the rotation curve has been varied as shown in Figure \ref{rvmodfig}b. Contour levels are as in Fig. \ref{bmhalo}.}
\label{rvpv}
\end{figure*}

Because the radial density profile and rotation curve are partially coupled in the PV diagrams, the density profile may not be responsible for the change in shape of the halo PV diagrams. We next attempt to modify the \emph{form} of the rotation curve such that the shape of the halo PV diagrams is better matched. In this case, the CR model radial density profile is unchanged. The 1 km s$^{-1}$ arcsec$^{-1}$ vertical gradient in azimuthal velocity and +10 km s$^{-1}$ systemic velocity offset are also still included. We have not attempted to recover the exact shape of the rotation curve in the halo. Rather, we illustrate how a change in the shape of the rotation curve changes the shape of the PV diagrams in the halo. The modification to the rotation curve is shown in Figure \ref{rvmodfig}b. The rotation speed rises roughly linearly for $R<50\arcsec$, as would be the case for a less centrally condensed potential. Figure \ref{rvpv} shows the PV diagram overlays. The modification to the radial density profile (\S \ref{rdenssection}) appears to match the data somewhat better than changing the shape of the rotation curve, but we cannot exclude the latter possibility.


\subsubsection{Modification of halo position angle and inclination}

\label{halopa}

In Figure \ref{mom}b, the appearance of the velocity contours in the SW and SE quadrants could be interpreted as resulting from the halo being oriented at a slightly different position angle relative to the disk. The effect is not clearly visible in the NE and NW quadrants, although the mean velocities in both spectra of \citet{rand00} are consistent with this asymmetry in all four quadrants. Such a shift in the position angle in a lagging halo would make the contours run more perpendicular to the major axis in two opposite quadrants (e.g., NW and SE in Fig. \ref{mom}b), and increase the angle between the contours and the minor axis in the other two quadrants (e.g., NE and SW in Fig. \ref{mom}b). Examination of the velocity contours indicates that the magnitude of such an offset in position angle necessary to produce the observed effect would be on the order of $20\degr$, but it should be noted that clumpiness in the halo gas distribution and peculiar velocities may confuse the situation.

On a similar note, if we allow the inclination of the halo to vary relative to that of the disk, it is possible that an apparent lag in halo rotation could be simply an indication that the halo is viewed from a more face-on perspective than the disk. If we assume a disk rotation speed of 200 km s$^{-1}$ and a vertical gradient in azimuthal velocity of 1 km s$^{-1}$ arcsec$^{-1}$ beginning at a height of $5\arcsec$, then at a height $z=30\arcsec$, the azimuthal velocities in the halo would be about 175 km s$^{-1}$. To mimic this effect with a variation in inclination angle, an offset between the disk and halo of about $30\degr$ would be required. We consider such an offset very unlikely. An offset in the position angle or inclination of the halo relative to the orientation of the disk could occur if the extraplanar gas were accreted. This possibility is interesting given the interaction between NGC 5775 and its neighbor, but it is uncertain how long such a configuration might last.

\section{The Ballistic Model}

\label{ballistic}

We next utilize the ballistic model of CBR to attempt to understand what vertical gradient in azimuthal velocity would be expected if the disk-halo flow in NGC 5775 is purely ballistic in nature. The full details of the model are described by CBR, who compared mean velocities from this model to those obtained from slit spectra for NGC 5775 and NGC 891. Basically, clouds are launched from the disk with an initial velocity selected from a constant probability distribution between zero and a maximum ``kick velocity'', $V_k$, along a unit vector normal to the midplane for models considered here. The clouds are then allowed to orbit ballistically in the galactic gravitational potential of \citet{wolf95}. Whenever a cloud leaves the simulation or returns to the disk, it is replaced by another cloud. Initial locations of the clouds are randomly selected from a distribution with a radial exponential scale length $R_0$ and a vertical gaussian scale height $z_0=0.2$ kpc. Other model parameters described by CBR are not used here. The clouds do not interact with each other, and are assumed to have constant temperature, density, and size (and therefore equal H$\alpha$ intensities) throughout the course of their orbits. After the simulation is run for 1 Gyr, at which time the system has reached a condition of steady state, positions and velocities of each cloud are extracted from the model. These outputs can be examined directly or used to generate an artificial data cube at any inclination, from which, e.g. PV diagrams or runs of mean velocity versus projected height above the midplane can be created.

\begin{deluxetable}{ll}
\tablecaption{Ballistic Base Model Characteristics for NGC 5775}
\tablehead{\colhead{Parameter} & \colhead{Value}}
\tablecolumns{2}
\tabletypesize{\scriptsize}
\startdata
$R_0$ & 6 kpc \\
$z_0$ & 200 pc \\
$V_k$ & 160 km s$^{-1}$ \\
$V_c$ & 198 km s$^{-1}$ \\
\enddata
\label{bmodtable}
\end{deluxetable}

The most influential parameter in the model is the ratio of the maximum kick velocity to the rotational velocity of the disk, $V_k/V_c$. For a given value of the circular velocity and initial $R$, increasing kick velocities result in more radial movement, larger maximum height of the orbit, and increased drop in azimuthal velocity at the peak height of the orbit. The circular velocity is determined observationally, and the kick velocity is set such that the resulting scale height of cloud density matches the scale height of EDIG emission (determined by CBR). This critical parameter is thus reasonably well constrained by observations. The other parameters were found to have little effect on the model outputs. The characteristics of the so-called base model for NGC 5775 from CBR are summarized in Table \ref{bmodtable}. With these parameters set in the model, CBR compared the mean velocities obtained from the model viewed at $i=86\degr$ with those measured along their two slits for NGC 5775. They found that the mean velocities from the model were roughly the same as the measured ones, but the model could not reproduce the observed mean velocities at the largest heights, which are seen to approach the systemic velocity (see their Figure 8).

\begin{figure}
\begin{center}
\epsscale{0.4}
\plotone{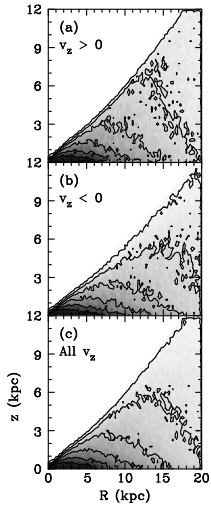}
\end{center}
\caption{Contour plots of cloud density as a function of galactocentric radius $R$ and $z$ in the ballistic ``base model'' (see text), for clouds (a) moving up, (b) moving down, and (c) all clouds. Contours levels correspond to 80, 400, 880, 1520, 2320, and 3280 clouds kpc$^{-2}$ in (a) and (b). In (c), contour levels correspond to 160, 800, 1760, 3040, 4640, and 6560 clouds kpc$^{-2}$.}
\label{bmrdensfig}
\end{figure}

\begin{figure}
\begin{center}
\epsscale{0.4}
\plotone{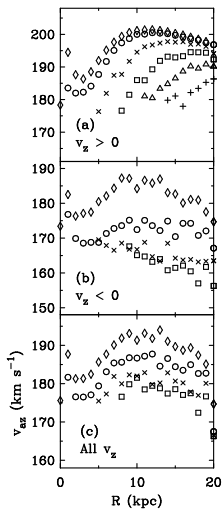}
\end{center}
\caption{Plots of average azimuthal velocity versus galactocentric radius $R$ in the ballistic base model, for clouds (a) moving up, (b) moving down, and (c) all clouds. Points are plotted in (a) for $z=0$ kpc (\emph{diamonds}), $z=2$ kpc (\emph{circles}), $z=4$ kpc (\emph{crosses}), $z=6$ kpc (\emph{squares}), $z=8$ kpc (\emph{triangles}), and $z=10$ kpc (\emph{plus signs}). The same symbols are plotted in (b) and (c), except that the $z=8$ kpc and $z=10$ kpc points are left out for clarity (the azimuthal velocities at those heights are very similar to the ones at $z=6$ kpc).}
\label{azvelfig}
\end{figure}

We have performed an analysis of the variation in azimuthal velocity as a function of $z$ in the ballistic model. This analysis shows that in the model that best matches the mean velocities from the slit spectra of CBR, the vertical gradient in azimuthal velocity is shallower than the corresponding variation in mean velocity. The reason for this discrepancy is the large-scale radial redistribution of clouds in the ballistic model. Figure \ref{bmrdensfig} displays contour plots of cloud density in the ballistic model as a function of $R$ and $z$, and Figure \ref{azvelfig} shows the azimuthal velocity curves of the ballistic model clouds as a function of height. Most of the clouds at high $z$ are also found at large $R$. Therefore, most of the decrease in mean velocity in the model is caused by velocity projection, and the magnitude of the vertical azimuthal velocity gradient is of lesser importance in setting the mean velocity gradient. For this reason, the results of the ballistic model cannot be relied upon to explain variations in mean velocity unless the observed radial density profiles are found to be similar to those in the model. In the present study, our spectral resolution is sufficient to allow the density distribution to be modeled, thus allowing azimuthal velocities to be measured as a function of $z$ and compared directly with azimuthal velocities from the ballistic model.


\begin{figure*}
\begin{center}
\epsscale{1}
\plottwo{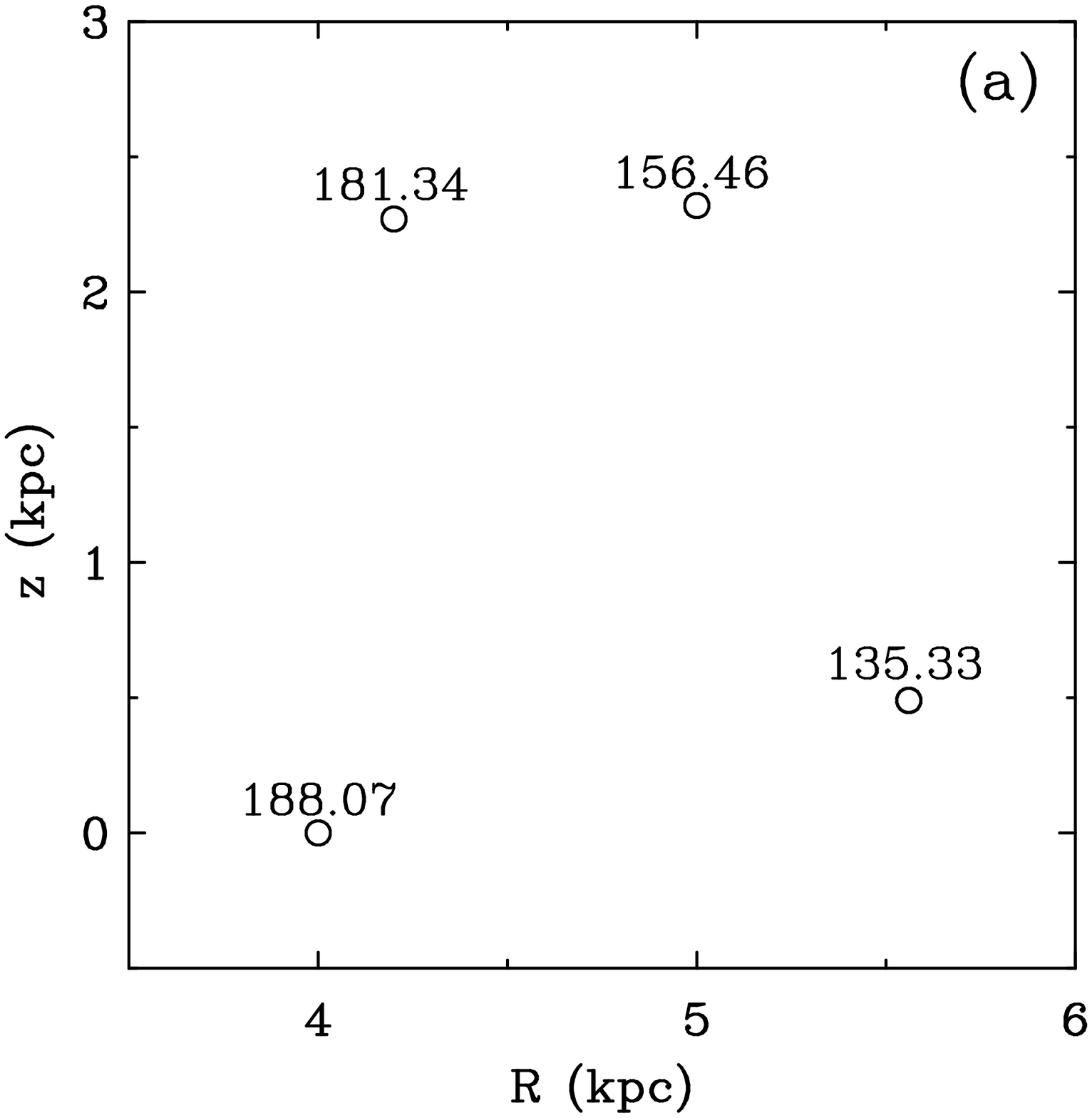}{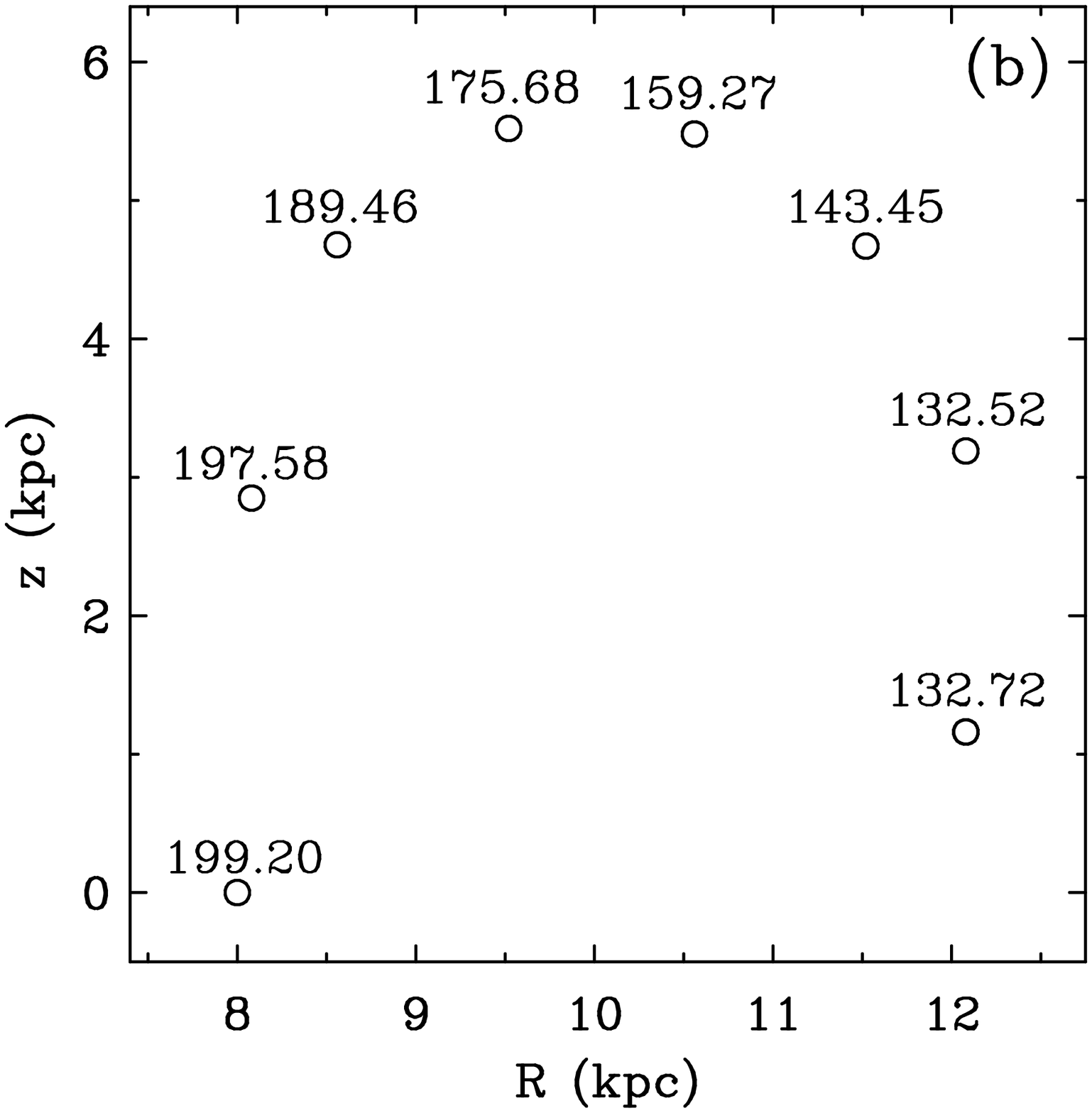}
\plottwo{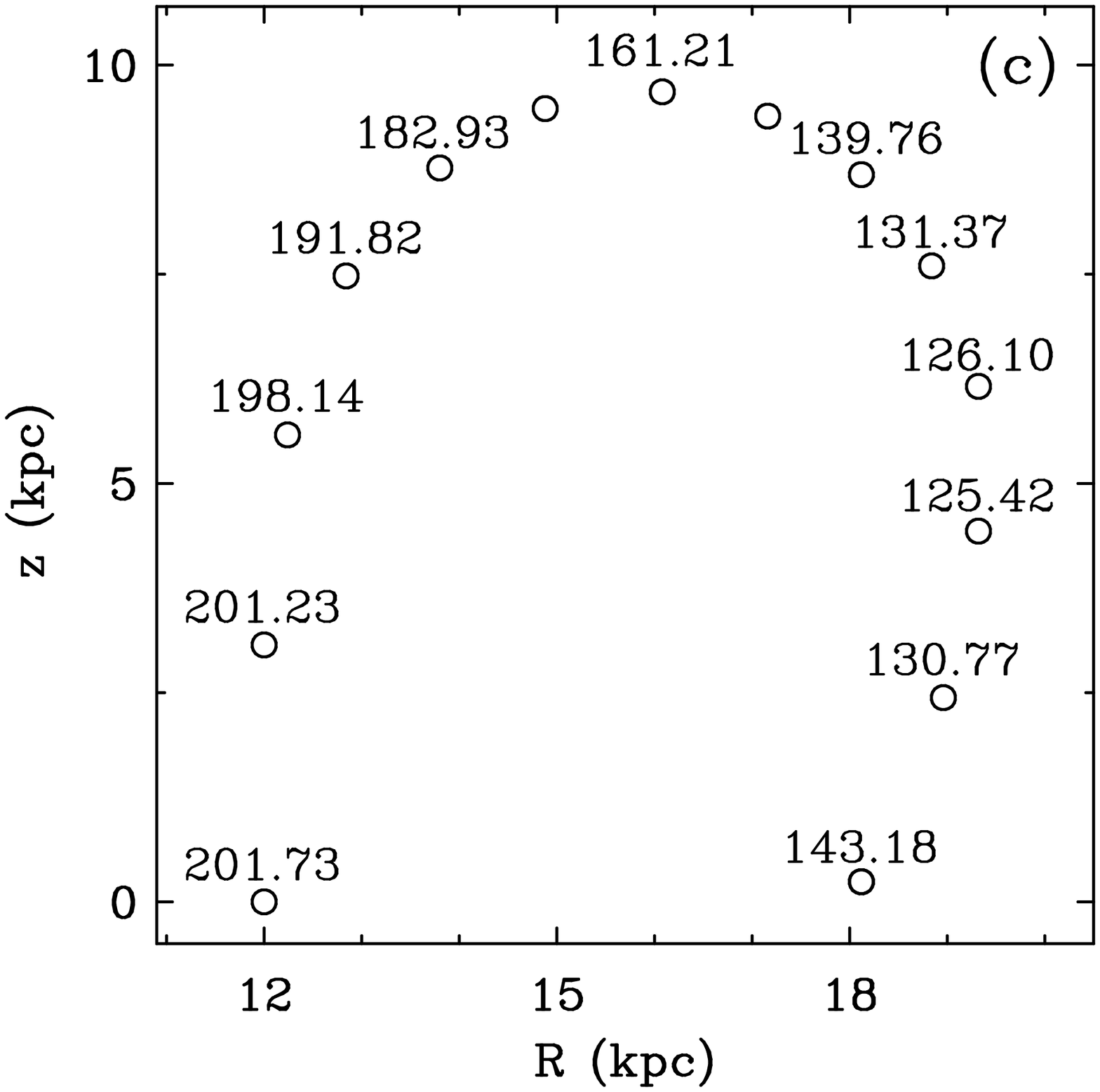}{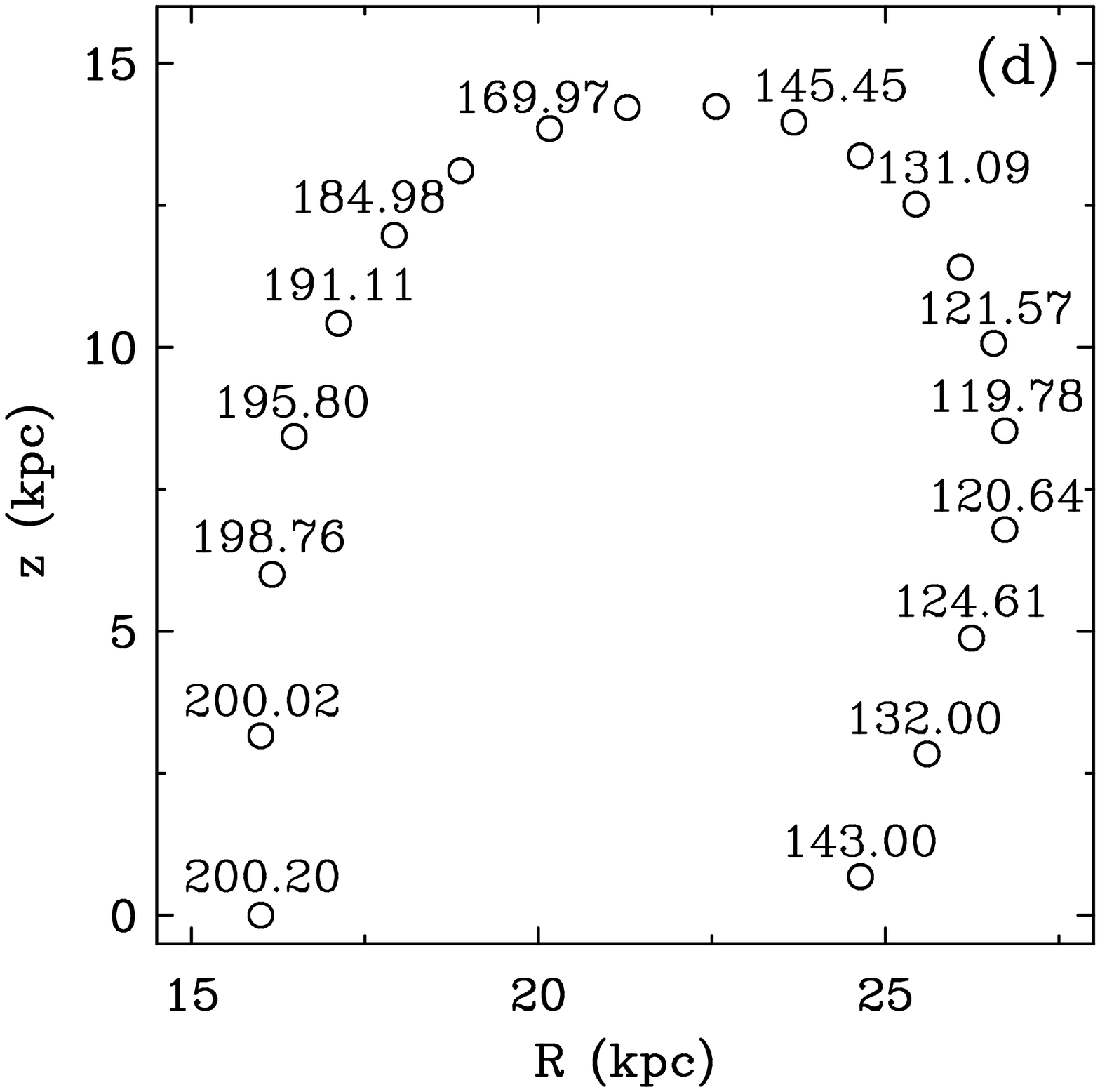}
\end{center}
\caption{Meridional plots of clouds in the ballistic base model with kick velocities equal to the maximum value, $V_k=160$ km s$^{-1}$, starting at galactocentric radii (a) $R=4$ kpc; (b) $R=8$ kpc; (c) $R=12$ kpc; and (d) $R=16$ kpc. Cloud positions are plotted at 20 Myr intervals. Azimuthal velocities (in km s$^{-1}$) are noted above the cloud's plotted position for most time steps (some omissions are made for clarity).}
\label{meridional}
\end{figure*}


\begin{figure}
\begin{center}
\epsscale{0.4}
\plotone{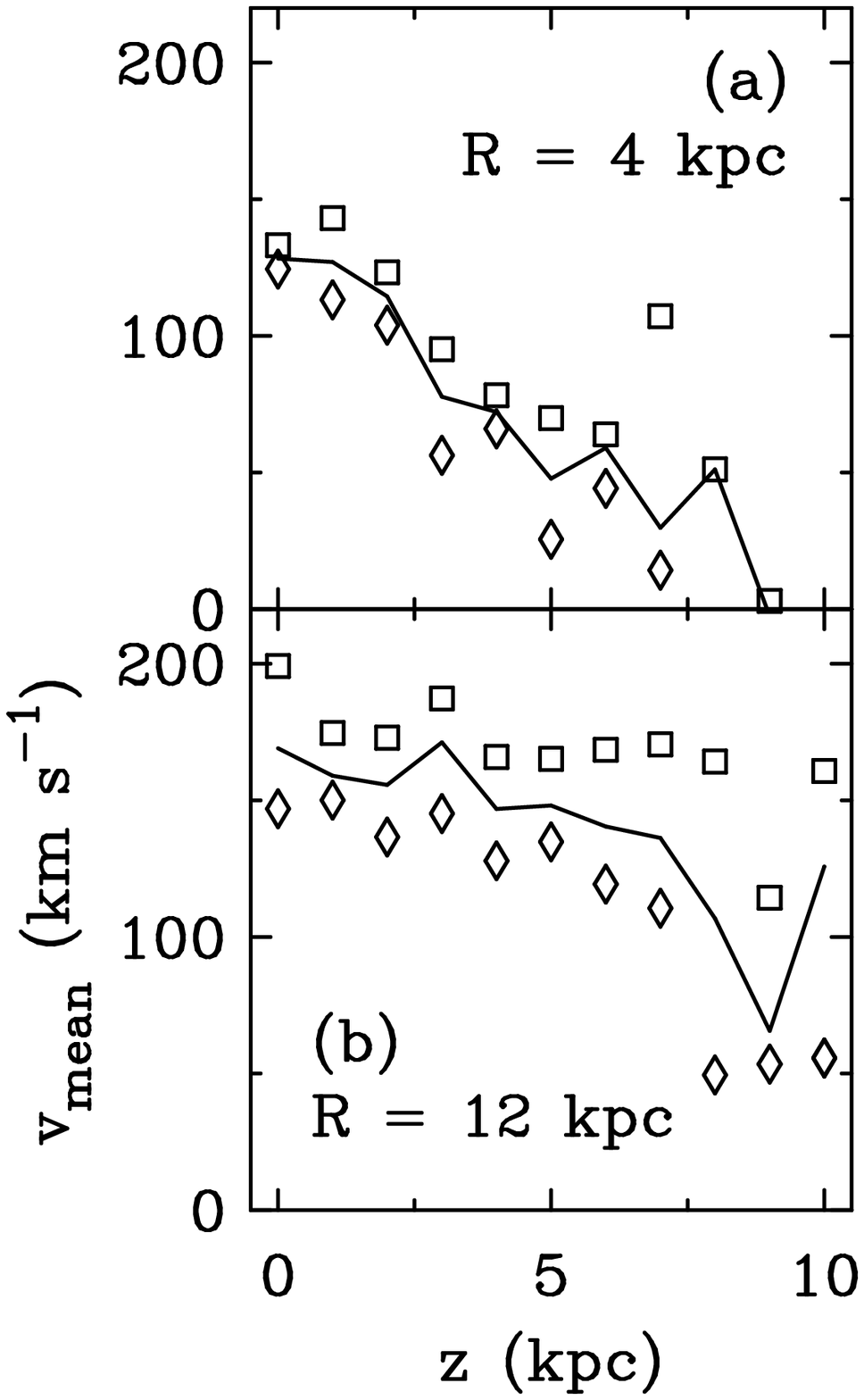}
\end{center}
\caption{Plots of mean velocity as a function of height above the midplane ($z$) in the ballistic base model for clouds at major axis distances (a) $R=4$ kpc and (b) $R=12$ kpc. Mean velocities are shown for clouds moving up (\emph{squares}), moving down (\emph{diamonds}), and all clouds (\emph{solid lines}). All mean velocities were calculated assuming an inclination angle of $86\degr$.}
\label{meanvelfig}
\end{figure}


As an aside, we point out how the azimuthal velocities to be extracted from the model may depend on the physical nature of the disk-halo flow. In Figure \ref{meridional}, we display meridional plots of cloud orbits in the ballistic model at various initial radii. The ballistic model predicts that the radial motion during the majority of a cloud's orbit is radially outward. Only at the highest initial radii and kick velocities do the ends of the orbits show a radially inward motion. Because the radius of a given cloud is nearly always increasing, conservation of angular momentum dictates that the azimuthal velocity of the same cloud is nearly always decreasing. The azimuthal velocities of upward-moving clouds are thus always higher than those of downward-moving clouds. The vertical gradient in azimuthal velocity extracted from the model will therefore be dependent on the assumption of the dynamics of the disk-halo flow. In a scenario where the gas begins in the ionized state, cools and condenses into neutral clouds somewhere near the top of the clouds' orbits, the azimuthal velocities will be relatively high. On the other hand, if the flow begins as hot gas and cools to a warm ionized gas for the downward portion of the flow, the azimuthal velocities will be lower. In the latter case, the assumption of ballistic motion is likely violated for the upward portion of the flow, and simply associating the warm ionized gas with the downward moving clouds in our model may not be accurate. The fact that the structure of the EDIG emission resembles shells and filaments implies that at least part of the ionized distribution is upward moving, but additional evidence is required to determine how much of the flow is ionized.

Clearly, the vertical gradient in azimuthal velocity in Figure \ref{azvelfig} is not as simple as that described by equation \ref{dvdzeqn}, which assumes a constant rotation curve shape as a function of height. Nevertheless, an approximate value of $dv/dz$ can be obtained by inspecting the plots. In the interest of comparison with the velocity gradient modeled from the data, we only report here the gradient seen in the ballistic model for 0 kpc $\leq z\leq$ 4 kpc, which is a slightly larger vertical range than we have modeled in this work (\S \ref{modeling}), and for 5 kpc $\leq R \leq$ 12 kpc. (No clouds in the model reach $z=4$ kpc for $R<5$ kpc, so azimuthal velocities cannot be extracted at those locations. The upper limit, 12 kpc, corresponds approximately to the optical radius of NGC 5775.) For the upward-moving clouds, the average gradient over the indicated ranges of $R$ and $z$ is approximately 2.4 km s$^{-1}$ kpc$^{-1}$; for the downward-moving clouds, 4.3 km s$^{-1}$ kpc$^{-1}$; for all clouds considered together, 2.6 km s$^{-1}$ kpc$^{-1}$. Even the highest value, 4.3 km s$^{-1}$ kpc$^{-1}$ (appropriate if all of the observed H$\alpha$ emission is from clouds returning to the disk), is approximately a factor of 2 lower than the vertical gradient modeled from the data [about 8 km s$^{-1}$ kpc$^{-1}$; recall that a galaxy model using $dv/dz \approx 4$ km s$^{-1}$ kpc$^{-1}$ was examined and rejected (cf. \S \ref{vvgsection})].

The base model of CBR includes a ratio of $V_k/V_c=0.81$ in order to match the observed scale height of EDIG emission. To understand how the maximum kick velocity affects the vertical gradient in azimuthal velocity, and thereby to examine whether a model with a different value of $V_k$ might better match the gradient estimated from the data, we have generated models with $V_k=100$ km s$^{-1}$ ($V_k/V_c=0.51$) and $V_k=220$ km s$^{-1}$ ($V_k/V_c=1.11$). In those cases, the gradients in azimuthal velocity were found to be approximately 1.9 km s$^{-1}$ kpc$^{-1}$ and 6.8 km s$^{-1}$ kpc$^{-1}$, respectively, for downward-moving clouds only. The approximately linear relationship thus determined between the gradient in azimuthal velocity and $V_k/V_c$ implies that a maximum kick velocity of 255 km s$^{-1}$ ($V_k/V_c=1.29$) is required to match the observed gradient even when only downward-moving clouds are considered. Unfortunately, this value of $V_k/V_c$ yields a DIG layer with a vertical exponential scale height of approximately 6.3 kpc, or about a factor of 3 higher than the values derived from the two slit spectra used by CBR of $2.1-2.2$ kpc. We conclude that it is unlikely that any value of $V_k$ will result in a model that reproduces the observed gradient in azimuthal velocity.

CBR also consider two galactic potential models (2 and 2i) from \citet{db98}. They find that model 2i, which has an oblate, flattened halo \citep[axial ratio $q=0.3$ rather than $q=0.8$ as in model 2; see][]{db98}, produces a steeper change in radial migration with the ratio $V_k/V_c$ compared to their base model. We might therefore expect a steeper gradient in azimuthal velocities with height if we use this potential. Referring to Figures 4 and 5 of CBR, we select $V_k=200$ km s$^{-1}$ for model 2i, which should give roughly the same radial redistribution as the base model with $V_k=255$ km s$^{-1}$, for radii $R \gtrsim 5$ kpc. The gradients in azimuthal velocity generated by this model are approximately 3.2 km s$^{-1}$ kpc$^{-1}$ for downward-moving clouds, 1.0 km s$^{-1}$ kpc$^{-1}$ for upward-moving clouds, and 1.6 km s$^{-1}$ kpc$^{-1}$ for all clouds considered together. This model produces a disk with an average vertical scale height of about 3.5 kpc (considerably higher than the observed scale height, $2.1-2.2$ kpc). Increasing the maximum kick velocity still higher to increase the azimuthal velocity gradient will only make the scale height larger. Thus, a flattened dark halo is not likely to be more successful in explaining the observed gradient in azimuthal velocity.

Figure \ref{meanvelfig} demonstrates how the gradient in mean velocity in the model is affected not only by the gradient in azimuthal velocity but also by the radial redistribution of clouds. The gradient in mean velocity versus height in the model is shown for $R=4$ kpc and $R=12$ kpc. Fig. \ref{meanvelfig}a ($R=4$ kpc) shows a much steeper gradient in mean velocity with height than Fig. \ref{meanvelfig}b ($R=12$ kpc). The average gradients over the range 0 kpc $\leq z \leq$ 4 kpc are approximately 14 km s$^{-1}$ kpc$^{-1}$ in the former case and 6 km s$^{-1}$ kpc$^{-1}$ in the latter, despite the gradient in azimuthal velocity being roughly the same at these two radii. In Fig. \ref{bmrdensfig}, the cloud distribution reaches $3-4$ times higher at $R=12$ kpc than at $R=4$ kpc. Thus, the only clouds encountered along a line of sight at low radius and high $z$ are concentrated at the edges of the disk, whereas a line of sight at higher radius and the same $z$ encounters more clouds closer to the line of nodes, keeping the mean velocity closer to the actual rotation speed at that height. At both radii, the gradient in mean velocity is much larger than that of azimuthal velocity because of radial redistribution.

\begin{figure}
\begin{center}
\includegraphics[angle=270,scale=0.4]{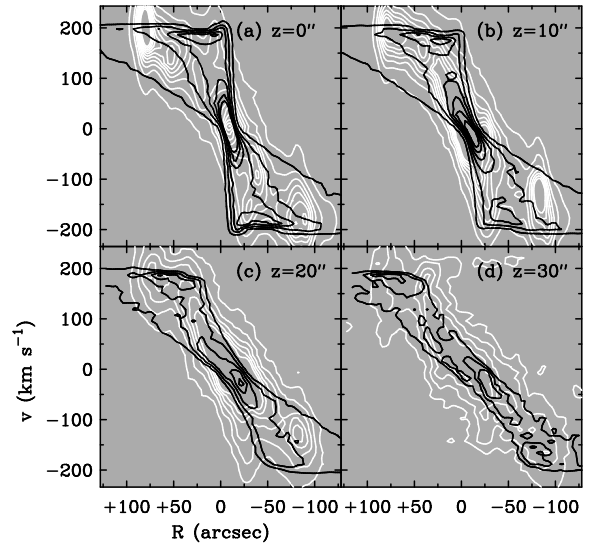}
\end{center}
\caption{PV diagrams constructed from the output of the ballistic base model (\emph{black contours}), which has been projected to the distance of NGC 5775, smoothed to an $8\arcsec$ beam, and viewed at $i=86\degr$. PV diagrams were created at heights of (a) $z=0\arcsec$, (b) $z=10\arcsec$, (c) $z=20\arcsec$, and (d) $z=30\arcsec$. Contour levels correspond to 0.5, 2, 3.5, 5, 6.5, and 8 clouds arcsec$^{-2}$ channel$^{-1}$ in (a); 0.5, 1.5, 2.5, 3.5, 4.5, and 5.5 clouds arcsec$^{-2}$ channel$^{-1}$ in (b); 0.5, 1, 1.5, 2, and 2.5 clouds arcsec$^{-2}$ channel$^{-1}$ in (c); 0.5, 0.75, and 1 clouds arcsec$^{-2}$ channel$^{-1}$ in (d). Also plotted in each frame are PV diagrams constructed from the H$\alpha$ data cube (\emph{white contours}) on the positive-$z$ (southwest) side of the halo. Contour levels correspond to 20$\sigma$ to 520$\sigma$ in increments of 50$\sigma$ in (a); 20$\sigma$ to 200$\sigma$ in increments of 30$\sigma$ in (b); 5$\sigma$ to 35$\sigma$ in increments of 6$\sigma$ in (c); 2$\sigma$ to 11$\sigma$ in increments of 3$\sigma$ in (d). The channel width in the ballistic model is the same as that of the H$\alpha$ data cube, 11.428 km s$^{-1}$. Positive values of the major axis distance $R$ correspond to the southeast side of the disk.}
\label{bmpvfig}
\end{figure}

\begin{figure}
\begin{center}
\epsscale{0.4}
\plotone{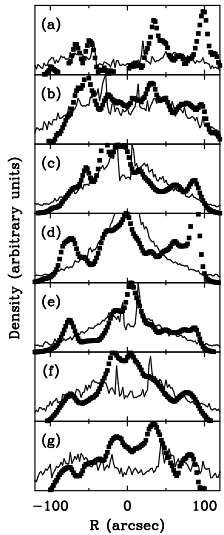}
\end{center}
\caption{Comparison between moment-0 cuts generated from the H$\alpha$ data (\emph{squares}) and the ballistic base model (\emph{solid lines}), at heights of (a) $z=-30\arcsec$, (b) $z=-20\arcsec$, (c) $z=-10\arcsec$, (d) $z=0\arcsec$, (e) $z=+10\arcsec$, (f) $z=+20\arcsec$, and (g) $z=+30\arcsec$. In each panel, the mean of the model profile is scaled to the mean of the data. The major axis distance $R$ is positive on the receding (southeast) side of the disk, and $z$ is positive to the southwest.}
\label{mom0cutfig}
\end{figure}

PV diagrams constructed from the output of the ballistic base model and displayed in Figure \ref{bmpvfig} clearly illustrate the points previously discussed and may be directly compared to the data. Along the major axis (Fig. \ref{bmpvfig}a), the appearance of the ``knee'' in the ballistic model PV diagram is roughly matched to that in the data. We note that the radial density profile and major axis rotation curve in the ballistic model have not been adjusted to the extent described in \S \ref{modeling}, and that the ballistic model does not simulate noise, so signal-to-noise contours may not be plotted and directly compared to the data. As in the observations, the ``knee'' in the ballistic model PV diagrams moves radially outward with increasing height above the midplane. Simultaneously, the velocity profile peak at a given $R$ shifts away from the local azimuthal velocity and closer to the systemic velocity. Both effects are caused primarily by the radial redistribution of clouds in the halo, and appear more striking in the ballistic model than in the data (see also Figure \ref{mom0cutfig}). The PV diagrams in Fig. \ref{bmpvfig} make clear that, at least in the ballistic model, most of the changes are caused by radial motions reshaping the velocity profiles. A gradient in azimuthal velocity is present in the model, and, though clearly evident by inspecting the shape of the PV diagrams, produces relatively minor modifications. That the effects described above are also seen in the observed PV diagrams provides further evidence for radial redistribution in NGC 5775. 


Whether the large-scale radial redistribution of gas predicted by the ballistic model is actually observed in NGC 5775 is a very important question. Such a redistribution may be apparent in cuts taken parallel to the major axis through moment-0 maps at various heights above the plane. Figure \ref{mom0cutfig} shows such cuts for moment-0 maps made from the base ballistic model and the H$\alpha$ data cube. At most heights, it appears that the data cuts may follow the shape of the model cuts fairly well, apart from the complication of filamentary structures. This correspondence is better for $z<0\arcsec$. At heights greater than $30\arcsec$, the moment-0 cuts from the ballistic model show a much clearer signature of radial redistribution, but our data are not sensitive enough to make a comparison at those heights. We conclude, then, that there is moderate morphological evidence for radial redistribution at the level predicted by the ballistic model.


CBR have previously compared the results of this ballistic model to mean velocities obtained with slit spectra for NGC 891 and NGC 5775. In NGC 891, the authors find that the observed mean velocities drop more slowly as a function of height than the mean velocities derived from the ballistic model. The authors concluded from this that (magneto-) hydrodynamical effects are probably at work. For instance, the radial migration could be modified by a gas pressure gradient in the halo. However if, as seems reasonable, halo gas pressure decreases with radius, the discrepancy would be worse because outward radial migration would be larger and modeled mean velocities even closer to systemic. It is possible that one major reason for the difference between mean velocities from the data and the model is that the radial density distribution in the halo of NGC 891 is more centrally concentrated than the prediction of the ballistic model. That outward radial migration occurs at some level in the halo is indicated by Figure 11 of \citet{rand97}. However, to properly examine halo rotation, azimuthal velocities must be estimated from data and the density distribution modeled, as in the present study. In a forthcoming paper, we attempt such an analysis from high spectral resolution data of the DIG halo of NGC 891 with two-dimensional spatial coverage.

In the case of NGC 5775, CBR find a better overall agreement between the mean velocities obtained from the slit data and those from the ballistic model. However, the ballistic model does not reproduce all of the observed features. Of particular interest with respect to the present study, the modeled mean velocities in the range $z=0-4$ kpc (the same range considered here) show a markedly steeper gradient than the data. Mean velocities calculated from the Fabry-Perot data cube show a similar lack of agreement with those from the ballistic model. In both cases, small-scale velocity and density variations are likely significantly affecting the observed mean velocities.

\section{Conclusions}

\label{conclusion}

We have obtained Fabry-Perot spectra of the H$\alpha$ emission line in the nearly edge-on spiral galaxy NGC 5775. Major axis radial density profiles and rotation curves were obtained for the ionized and molecular components of this galaxy. The observations have also allowed us to examine the azimuthal velocity variation as a function of height in the halo. We have found that a vertical gradient in the azimuthal velocity with a magnitude 1 km s$^{-1}$ arcsec$^{-1}$, or about 8 km s$^{-1}$ kpc$^{-1}$, is able to reproduce the gross features of PV diagrams constructed parallel to the major axis (but a larger gradient may be appropriate in localized regions of the halo). Such a gradient is primarily indicated for the approaching side of the galaxy, though mean velocities from slit 2 of \citet{rand00} suggest the presence of a gradient on the receding side as well (at least at the highest $z$, but recall that the radial density profile was not taken into account in that analysis). The magnitude of the gradient should be considered approximate because of uncertainties in the radial density profile and rotation curve adopted (especially for the halo gas) in this study. One should also be aware that the interaction with NGC 5774 may lead to significant deviations from axisymmetry in the halo, as has been assumed in the modeling here. Nevertheless, it is apparent that a non-zero gradient is present in this galaxy. Comparisons between PV diagrams constructed from the data and from galaxy models suggest either a radial redistribution of gas in the halo or a shallower rise in the rotation curve than is observed along the major axis. Further evidence for the former possibility is provided by total intensity profiles constructed at various heights parallel to the major axis.

The ballistic model of CBR has been analyzed in more detail. We have found that while azimuthal velocities decrease gently with increasing $z$, the corresponding mean velocities decrease more steeply (particularly at lower radius). This steep gradient in mean velocity is due to radial outflow of gas in the ballistic model. This result emphasizes the importance of using caution when interpreting mean velocities in edge-on or nearly edge-on galaxies. If the radial density profile of radiating gas is not well understood, the use of mean velocities as indicators of rotation speed can be highly misleading.

The decrease in azimuthal velocity with height in the ballistic model is, in fact, shallower than that which has been inferred from the data, suggesting that additional mechanisms are important. We note here some effects which may be at work in the halo of NGC 5775. In a picture of a fluid disk in hydrostatic equilibrium, as described by, e.g., \citet{b00}, the steeper gradient inferred from the data may imply pressure declining with radius in the halo, which is not unreasonable to expect. By considering the baroclinic solutions to stationary hydrodynamics, \citet{barn04} were successful in generating a model in agreement with the vertical gradient in the rotation curve of NGC 891 \citep{frat04c}; it would be interesting to test whether this method is able to reproduce the gradient measured in this work as well. A completely different physical picture which must also be considered is that of gas accretion. \citet{kauf05} were also able to reproduce the lag in the halo of NGC 891 with SPH simulations of infalling multiphase gas. Although we have considered the halo in this actively star forming galaxy to be star formation driven, the fact that NGC 5775 is interacting with its companion may suggest that such considerations are relevant here.

\acknowledgments{We thank Judith Irwin for the H$\,$I data, Siow-Wang Lee for the CO data, Ralph T\"ullmann for his slit spectrum, and Filippo Fraternali for the use of his GIPSY code. We also thank an anonymous referee for a detailed report that has led to significant improvement of our paper. This material is based on work partially supported by the National Science Foundation under Grant No. AST 99-86113. This research has made use of NASA's Astrophysics Data System Bibliographic Services.}

\appendix

\section{Sky subtraction and wavelength solution determination for Fabry-Perot observations}

\label{fpcorr}

\begin{figure*}
\begin{center}
\epsscale{1}
\plottwo{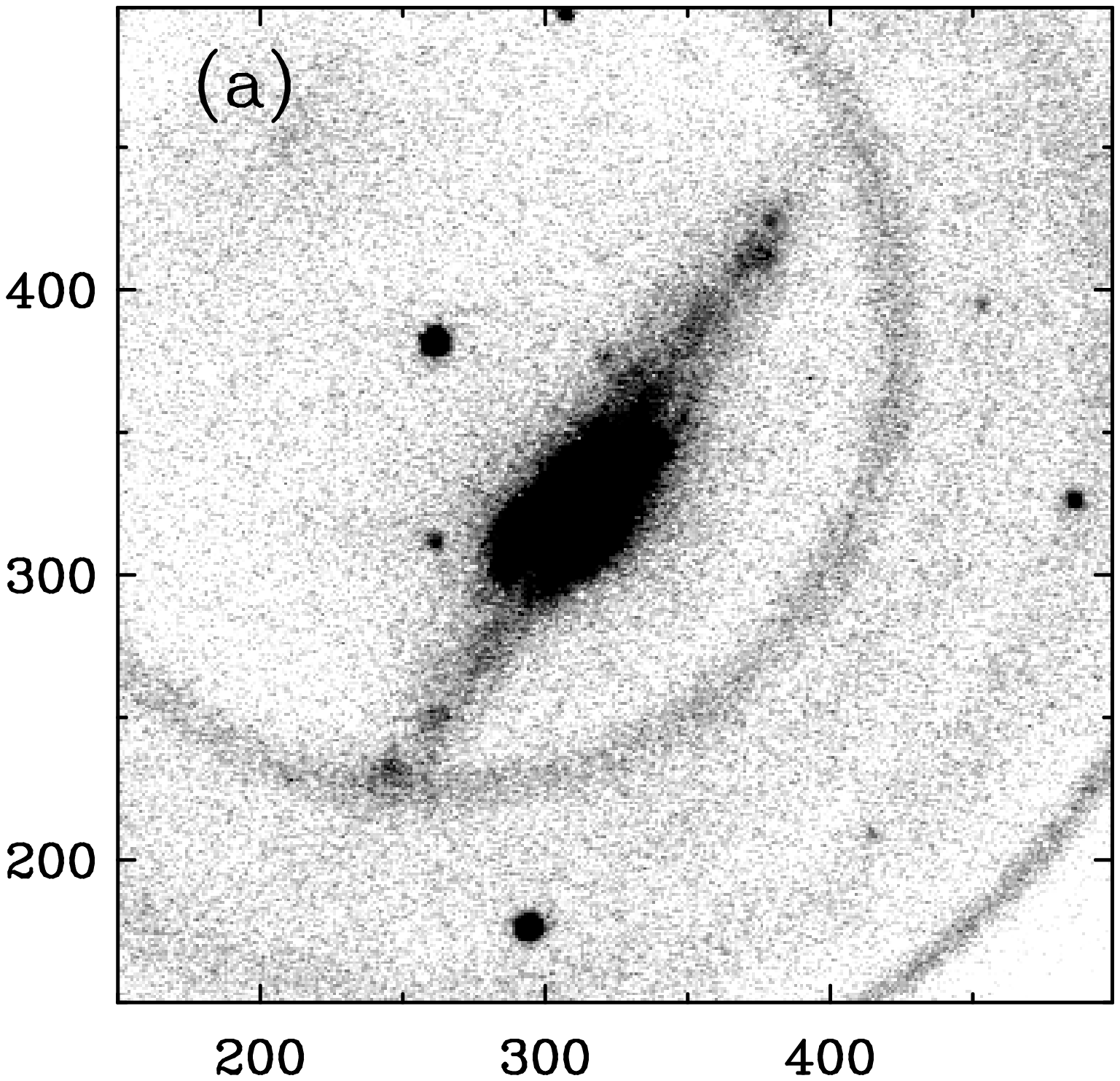}{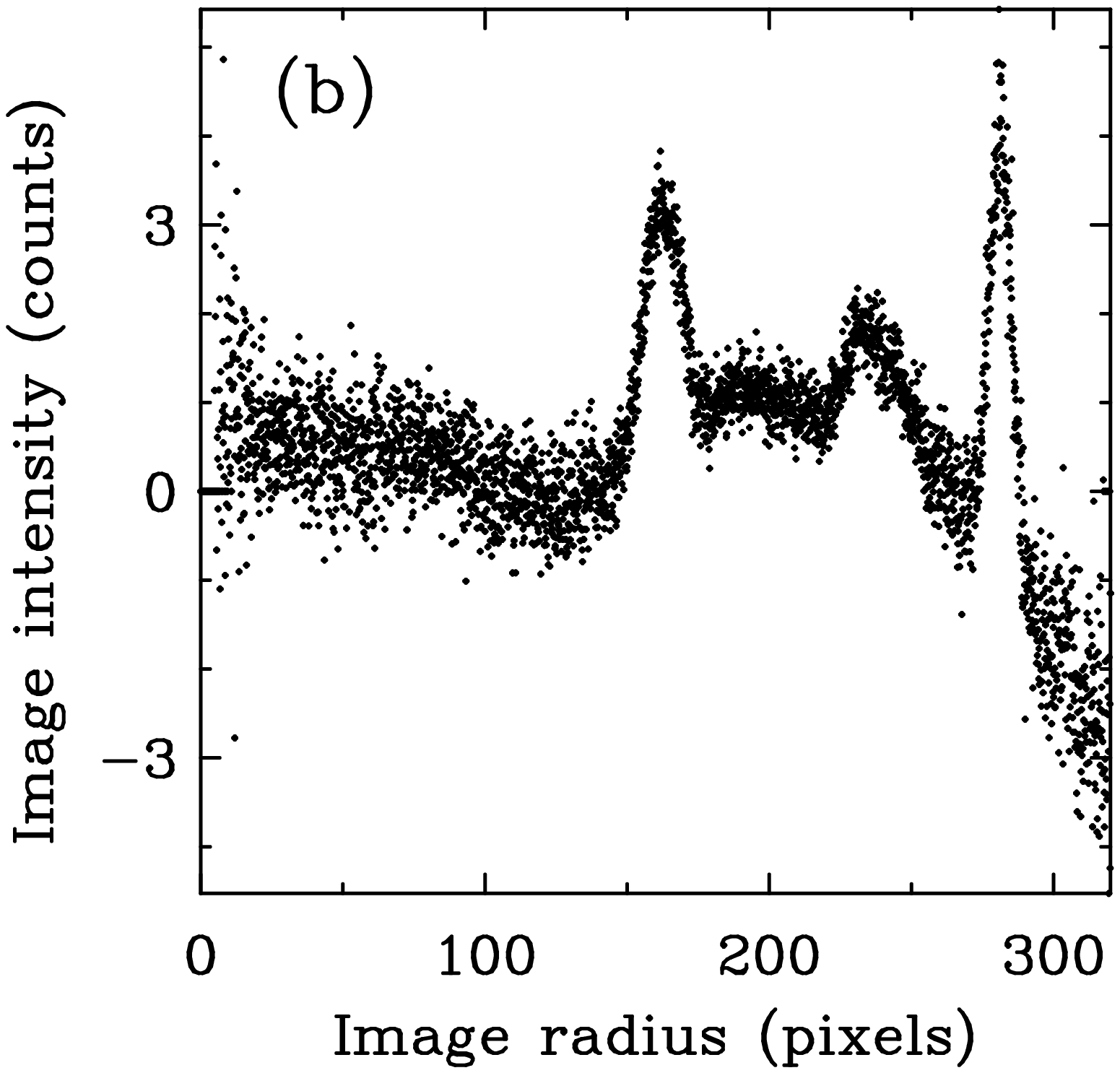}
\plottwo{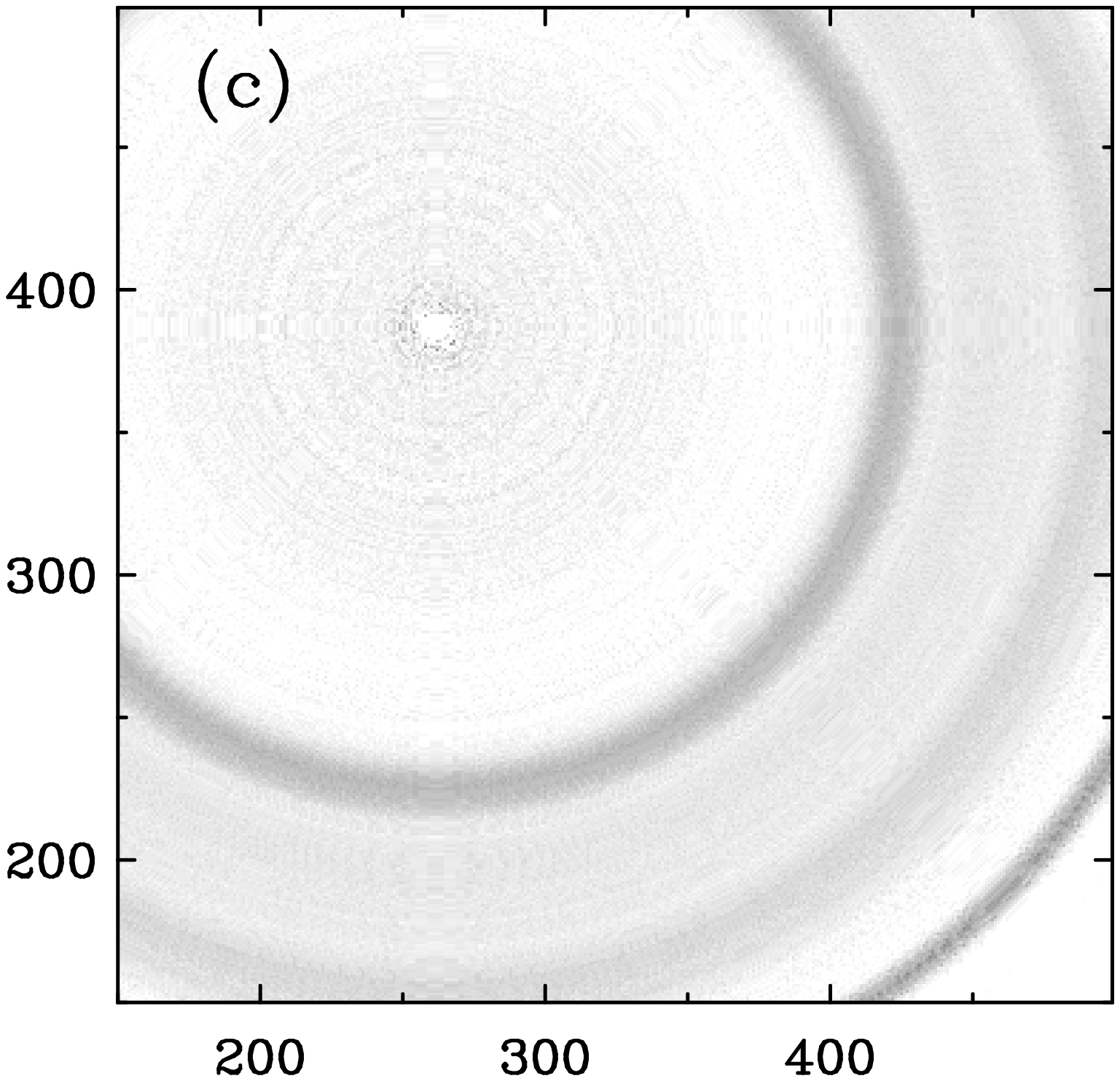}{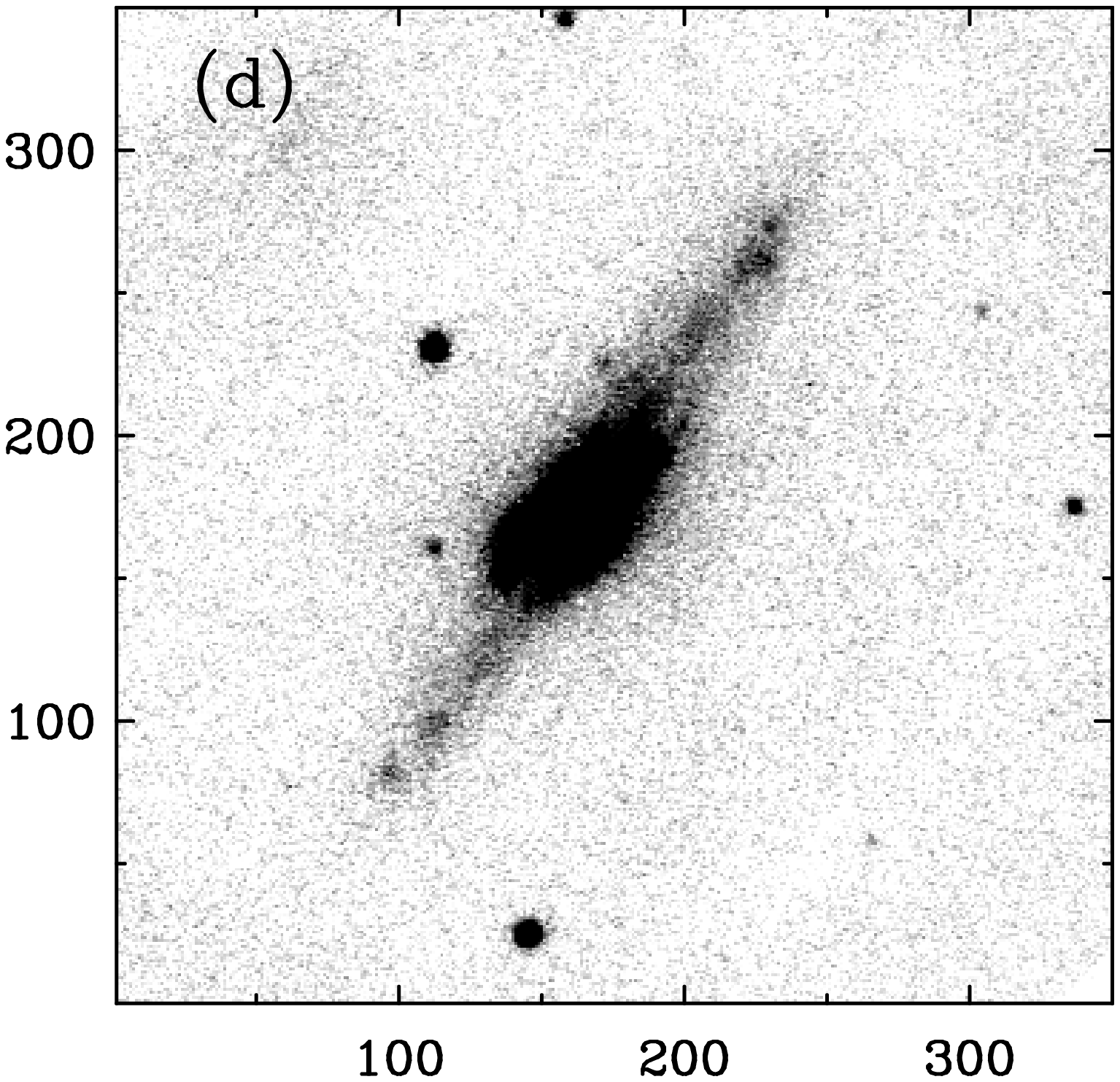}
\end{center}
\caption{(a) An original image at an individual etalon spacing. Note the prominent sky rings, and the faint ghost image in the upper-left corner. (b) Azimuthal average of the image displayed in \it a\rm, with the object pixels masked out. (c) The azimuthal average has been used to generate an image of the sky rings. (d) To generate this image, the image displayed in \it c \rm has been subtracted from that in \it a\rm. The same grayscale values have been used to display each image.}
\label{skysub}
\end{figure*}

The process of converting a Fabry-Perot data cube containing planes of constant etalon spacing ($h$) to a cube with planes of constant wavelength is called the phase correction. To complete the phase correction, the sky coordinates and wavelength of each pixel in the original cube must be specified. In this Appendix, we describe our method for subtracting the night sky line emission from the data, and determining the wavelength solution (the bulk of the phase correction process). Typically, arc-lamp comparison spectra are observed to facilitate obtaining the wavelength solution. Unfortunately, an arc-lamp cube was not obtained during our observing run. Therefore, night-sky emission lines were used for the wavelength calibration. Because the wavelength satisfying the etalon interference condition varies radially across each image, the night-sky emission lines appear as rings in the images (see Fig. \ref{skysub}a), with radii dictated by the wavelength of the line, the etalon spacing, and the curvature constant $K_{\lambda}$ (cf. equation \ref{instparms}). Thus, by measuring the properties of a ring (center and radius) in each frame so that it may be subtracted, the wavelength solution is also obtained. A first attempt at finding the optical center of each image was made by fitting a circle to three points on an individual ring, as described by \citet{bt89}. However, the rings appear at about the 3.5$\sigma$ level in the images, and because the widths of the ring profiles increase with decreasing ring radius it was often difficult to select appropriate points for a good fit. As noted by \citet{jones02}, the azimuthal symmetry of the ring images may be exploited to find the optical center and subtract the sky rings. The location of the optical center in each plane was found with the following algorithm:

\begin{enumerate}
\item First, the location of the optical center is approximated by fitting a circle to three points on an individual ring. This step yields a point close to the true optical center.
\item Next, object signal is masked, and an azimuthal average is performed about the fiducial center selected in step (1). This results in a radial profile with clear ring signatures. An example is shown in Fig. \ref{skysub}b.
\item Next, a non-linear least-squares algorithm is used to fit a gaussian to the most prominent ring profile.
\item The fiducial center point is then varied over a search grid. At each grid point, a new azimuthal average is calculated, and a gaussian is fit to the new ring profile. At the end of this step, the width of the line profile has been calculated at each fiducial optical center location.
\item Because of the azimuthal symmetry, the profile width is minimized at the true optical center. The grid of line widths obtained in step (4) is therefore interpolated, and the location of the minimum on the interpolated grid is taken to be the true optical center.
\end{enumerate}

The azimuthal profile corresponding to our choice of optical center is used to form a two-dimensional image of the sky rings (Fig. \ref{skysub}c), which is then subtracted from the original, resulting in a sky-subtracted image (Fig. \ref{skysub}d). Azimuthal variations in ring intensity and any non-circularity of the rings will result in incomplete subtraction of the sky emission. The residual emission was minimized by limiting the image area to a 4.6 arcminute square, inside of which the azimuthal variations were greatly reduced. The calculation of the azimuthal average was limited to this area in each step, and the sky-subtracted image was cropped to the same region.


The ring fitting algorithm also yields the radii of the sky rings in each image. The two brightest sky lines were identified as OH $\lambda 6596.64$ and the unresolved pair OH $\lambda\lambda 6603.99,6604.28$ \citep[cf.][]{osterbrock96}. We used a central wavelength of 6604.12\AA\ for the unresolved pair. For each image, the sky-line wavelengths $\lambda$ and radii $r$ are inserted in
\begin{equation}
\lambda = \lambda_0 + K_{\lambda}r^2
\label{klambda}
\end{equation}
to determine $\lambda_0$, the wavelength at the optical center. The curvature constant, $K_{\lambda}$, is dependent solely on instrumental parameters:
\begin{equation}
K_{\lambda}=(n/2)\Delta\lambda_0(p_{\mu}^2/f_{\mathrm{cam}}^2),
\label{instparms}
\end{equation}
where $n$ is the etalon order, $\Delta\lambda_0$ is the on-axis free spectral range, $p_{\mu}$ is the CCD pixel size, and $f_{\mathrm{cam}}$ is the focal length of the camera lens \citep{bt89}. Once $\lambda_0$ is known for each image, equation \ref{klambda} can be used to determine the value of $\lambda$ at every pixel (using its distance, $r$, from the optical center). The value of $K_{\lambda}$ [actually, its counterpart expressed in terms of etalon spacing, $K_h=(n/2)\Delta h_0 (p_{\mu}^2/f_{\mathrm{cam}}^2)$], was verified by inserting the nominal values for the instrumental parameters and comparing to the curvature value obtained by fitting a simple parabola to a plot of night-sky emission ring radius versus etalon spacing. The values of $K_h$ obtained with these two methods were found to differ by only 0.2 per cent; therefore, the nominal instrumental parameters were used to specify $K_{\lambda}$.

Once the value of $\lambda$ is known for each pixel, and the astrometric solutions have been obtained for each plane of the data cube (cf. \S \ref{observations}), the phase correction may be completed.

\end{document}